\pdfoutput=1
\documentclass[prd,twocolumn,nofootinbib,superscriptaddress]{revtex4-2}
\usepackage{graphicx}
\usepackage{bm}
\usepackage[colorlinks=true,unicode]{hyperref}
\usepackage{float}
\usepackage{amsmath}
\usepackage{amssymb}
\usepackage{pifont}
\usepackage{textcomp}
\usepackage{gensymb}
\usepackage{multirow}
\usepackage[normalem]{ulem}
\usepackage[dvipsnames]{xcolor}
\usepackage[utf8]{inputenc}

\newcommand{\kent}[1]{{#1}}

\newcommand{\sid}[1]{{#1}}
\newcommand{\bgmetric}{{}^0 g}
\newcommand{\pertmetric}{\delta g}
\newcommand\be{\begin{equation}}
\newcommand\ba{\begin{eqnarray}}
\newcommand\ee{\end{equation}}
\newcommand\ea{\end{eqnarray}}
\newcommand\bw{\begin{widetext}}
\newcommand\ew{\end{widetext}}
\newcommand\abzero{(\alpha,\beta) \to 0}

\begin{document}
\title{I-Love-Q in Ho\v rava-Lifshitz Gravity }
\author{Siddarth Ajith}
\affiliation{Department of Physics, University of Virginia, Charlottesville, Virginia 22904, USA}
\author{Kent Yagi}
\affiliation{Department of Physics, University of Virginia, Charlottesville, Virginia 22904, USA}
\author{Nicol\'as Yunes}
\affiliation{Department of Physics, University of Illinois at Urbana-Champaign,
Urbana, Illinois 61801, USA}

\date{\today}
\begin{abstract}
   Ho\v rava-Lifshitz  gravity is an alternative theory to general relativity which breaks Lorentz invariance in order to achieve an ultraviolet complete and power-counting renormalizable theory of gravity. In the low-energy limit, Ho\v rava-Lifshitz gravity coincides with a vector-tensor theory known as khronometric gravity. The deviation of khronometric gravity from general relativity can be parametrized by three coupling constants: $\alpha$, $\beta$, and $\lambda$. Solar system experiments and gravitational wave observations impose stringent bounds on $\alpha$ and $\beta$, while $\lambda$ is still relatively unconstrained ($\lambda\lesssim 0.01$). In this paper, we study whether one can constrain this remaining parameter with neutron star observations through the universal I-Love-Q relations between the moment of inertia (I), the tidal Love number (Love), and the quadrupole moment (Q), which are insensitive to details in the nuclear matter equation of state.
   To do so, we perturbatively construct slowly-rotating and weakly tidally-deformed neutron stars in khronometric gravity. We find that the I-Love-Q relations are independent of $\lambda$ in the limit $\abzero$. Although some components of the field equations depend on $\lambda$, we show through induction and a post-Minkowskian analysis that slowly-rotating neutron stars do not depend on $\lambda$ at all. Tidally deformed neutron stars, on the other hand, are modified in khronometric gravity (though the usual Love number is not modified, as mentioned earlier), and there are potentially new, non-GR Love numbers, though their observability is unclear. These findings indicate that it may be difficult to constrain $\lambda$ with rotating/tidally-deformed neutron stars. 
\end{abstract}
\maketitle

\section{Introduction}

General relativity (GR) is the current benchmark for gravitational theories, and it has continually passed all tests to date~\cite{Will:2018bme,Will:2014kxa,Berti:2015itd}.
However, it is still worthwhile to develop new theories to test GR and develop a theory of gravity beyond GR. Current interests include finding a valid theory of quantum gravity as well as explaining cosmological phenomena such as dark energy and dark matter~\cite{Jain:2010ka,Clifton:2011jh,Joyce:2014kja,Koyama:2015vza}. The problem with the former is that GR is not power-counting renormalizable.

Ho\v rava proposed a theory of gravity beyond GR that is renormalizable and ultraviolet complete~\cite{Horava:2009uw}. The theory breaks Lorentz invariance in the ultraviolet regime by introducing a Lifshitz-type anisotropic scaling between space and time (and thus the theory is called Ho\v rava-Lifshitz gravity). In the low-energy limit, the theory coincides with khronometric gravity~\cite{Blas:2010hb}. The latter contains a khronon scalar field whose constant hypersurfaces provide a time foliation of spacetime. Thus, the gradient of the khronon indicates a preferred direction in spacetime that violates Lorentz invariance, and khronometric gravity is an example of a vector-tensor theory. Though Lorentz violation is heavily constrained in the matter sector~\cite{Kostelecky:2003fs,Mattingly:2005re,Jacobson:2005bg,Liberati:2013xla}, it is not as stringently constrained in the gravity sector~\cite{Yagi:2013ava,Yagi:2013qpa,Gumrukcuoglu:2017ijh}. Khronometric gravity corresponds to a broader vector-tensor theory called Einstein-\AE ther theory~\cite{Jacobson:2000xp,Jacobson:2007veq} whose vector field (\ae ther field) is hypersurface orthogonal~\cite{Jacobson:2013xta}. 
Other vector-tensor theories include
many classes of general Proca~\cite{Heisenberg:2014rta,Allys:2015sht,Allys:2016jaq,Rodriguez:2017ckc,DeFelice:2016cri,GallegoCadavid:2020dho,Gomez:2019tbj,Nakamura:2018oyy} and beyond  general Proca theories~\cite{Heisenberg:2016eld,GallegoCadavid:2019zke}. 

Khronometric gravity is characterized by three coupling constants, $(\alpha,\beta,\lambda)$. Previous work on testing the theory has stringently constrained two of these three coupling parameters. From the comparison of gravitational waves from GW170817 and its associated gamma-ray burst GRB170817A~\cite{LIGOScientific:2017vwq}, along with considerations of parameterized post-Newtonian constraints from solar system experiments~\cite{Will:2014kxa,Blas:2011zd}, one can find a stringent constraint on $\alpha$ and $\beta$~\cite{Gumrukcuoglu:2017ijh,Ramos:2018oku}. This, along with theoretical considerations of stability~\cite{Blas:2009qj}, helium abundance from Big Bang nucleosynthesis~\cite{Chen:2000xxa,Carroll:2004ai,Gumrukcuoglu:2017ijh}, and cosmological constraints~\cite{Afshordi:2009tt}, imposes a constraint on the remaining parameter $\lambda$; the latter constraint, however, is still relatively weak in comparison to the constraint on the other two constants. There has been recent work to test this remaining parameter~\cite{Barausse:2019yuk,Franchini:2021bpt}, but no new bounds have been found beyond those mentioned above. As such, the focus of this work is to study whether tests of this remaining parameter can be carried out with the I-Love-Q universal relations of neutron stars~\cite{Yagi:2013awa,Yagi:2013bca}. 

Due to their compactness, neutron stars 
are excellent testbeds to probe strong-field gravity, and I-Love-Q is an excellent framework to do so~\cite{Yagi:2013awa,Yagi:2013bca,Yagi:2016bkt,Doneva:2017jop,Gupta:2017vsl}. The latter refers to universal relations between the moment of inertia (I), the tidal Love number (Love) and the quadrupole moment (Q) that are insensitive to the nuclear matter equation of state. The Love number has been constrained with GW170817~\cite{LIGOScientific:2017vwq,LIGOScientific:2018hze,LIGOScientific:2018cki} while the moment of inertia has been constrained with NICER observations~\cite{Silva:2020acr}, and it is also expected to be measured with the double pulsar binary PSR J0737-3039~\cite{Lattimer:2004nj,Kramer:2009zza,Hu:2020ubl}. Since the relations depend on the underlying gravitational theory and deviate from the GR ones as one considers non-GR theories, they allow us to study gravity beyond the limits of our current understanding of nuclear physics~\cite{Yagi:2013awa}. As an example of this, Silva \textit{et al}.~\cite{Silva:2020acr} recently combined the gravitational-wave constraints of the neutron star Love number using LIGO/Virgo data~\cite{LIGOScientific:2017vwq,LIGOScientific:2018hze,LIGOScientific:2018cki} with the X-ray constraints of the stellar compactness using NICER data~\cite{Miller:2019cac,Riley:2019yda}, together with the universal relation between the moment of inertia and compactness, to carry out the first multi-messenger tests of gravity with the universal I-Love relations. A follow-up analysis has been recently carried out using the universal relation between the tidal deformability and compactness~\cite{Saffer:2021gak}.

In this paper, we study whether one can use such a test to constrain khronometric gravity. Previous works have shown that asymptotically-flat, physical metrics in khronometric gravity coincide with GR \textit{in vacuum}~\cite{Bellorin:2010je,Franchini:2021bpt}, but to apply the multi-messenger test, we must study nonvacuum spacetimes, such as neutron stars. This paper therefore studies the structure of slowly-rotating and tidally-perturbed neutron stars in khronometric gravity, focusing in particular, on how the moment of inertia, the Love number and the quadrupole moment depend on $\lambda$. To achieve this, we follow the same procedure as in GR to construct slowly-rotating or weakly tidally-deformed neutron stars perturbatively in rotation and tidal deformation. We then use these solutions to extract the moment of inertia, the Love number and the quadrupole moment from the asymptotic behavior of the metric at a large distance from the star~\cite{Yagi:2013bca}. The analysis is similar to finding slowly-moving neutron stars in khronometric gravity to extract the stellar sensitivities~\cite{Yagi:2013qpa,Yagi:2013ava}. We also study other components of the metric perturbations, which are not needed to extract the I-Love-Q relations, but that do determine whether neutron stars in khronometric gravity are the same as in GR. We do so through a post-Minkowskian (PM) analysis in which we assume that the stellar compactness is small and solve the field equations order by order in compactness. A similar PM analysis has recently been used to find neutron star sensitivities in Einstein-\AE ther theory~\cite{Gupta:2021vdj} and scalar-tensor theories~\cite{Yagi:2021loe}.

Our main findings are as follows. First, we find that the field equations relevant for extracting the I-Love-Q relations are completely independent of $\lambda$ in the $\abzero$ limit. Thus, the relations are the same as in GR, which implies they cannot be used to constrain $\lambda$ with an I-Love-Q test. Second, we find that some components of the field equations do contain $\lambda$-dependence, although they do not affect the moment of inertia, the Love number or the quadrupole moment. Third, through the method of induction and a PM analysis, we show that even the solutions to these components of the field equations in the case of slowly-rotating neutron stars do not present khronometric gravity modifications, and thus, such stars in this theory are identical to those in GR when $\abzero$. On the other hand, tidally-deformed neutron stars acquire khronometric corrections that depend on $\lambda$. These corrections lead to new ``species'' of Love numbers that are absent in GR: two types are from the vector field of khronometric gravity (which we call ``vector Love numbers''), while the third type is from perturbations to the shift in the metric tensor (which we call ``shift Love number"). The observability of these new Love numbers is not yet clear. 

The rest of the paper is organized as follows. In Sec.~\ref{sec:field_eq}, we summarize the details of khronometric gravity and introduce the vector field as well as the field equations. We also discuss previously-found bounds on the theory. In Sec.~\ref{sec:metric_vector_matter_perturbations}, we summarize the ansatz for the spacetime metric, the vector field, and the matter stress-energy tensor used throughout the rest of the paper. In Sec.~\ref{sec:i_love_q}, we present our results in analyzing the I-Love-Q trio in khronometric gravity and show that the values coincide with those of GR in the limit $\abzero$. In Sec.~\ref{sec:V_H1_analysis}, we focus on the remaining components of the metric perturbations that are irrelevant to extracting the I-Love-Q trio and analyze them within a PM framework. In particular, we here derive new shift and vector\footnote{We note that Ref.~\cite{Pereniguez:2021xcj} finds a Love number which is called a ``vector Love number," but they are referencing the vector harmonic sectors of metric perturbations, as opposed to the vector field perturbations considered in this paper. } Love numbers for khronometric gravity. In Sec.~\ref{sec:conclusion}, we conclude and discuss future directions. In Appendix~\ref{app:scalarTensorAnalysis}, we discuss how khronometric gravity can be regarded either as a scalar-tensor or vector-tensor theory, and we show that either viewpoint leads to the same conclusion \kent{in terms of perturbations}. In Appendix~\ref{sec:AppendixUncontrainedValues}, we present the full field equations for neutron stars, keeping all the coupling constants.Throughout this work, we use the metric signature $(+,-,-,-)$, and we use geometric units $c=1=G_N$, where $G_N$ is the local Newtonian gravitational constant.

\section{Khronometric Gravity}
\label{sec:field_eq}

In this section we present the action and equations of motion for khronometric gravity. We also discuss some previous bounds on the theory, which indicate what values we consider for the coupling constants that parametrize the theory. 

First, we present the khronometric action, which is given by~\cite{Blas:2010hb,Barausse:2019yuk} 
\be\label{eq:action}
S=\frac{-1}{16\pi G_{\rm bare} \color{black}}\int d^4x\sqrt{-g}(R+\mathcal L)+S^\mathrm{(mat)},
\ee
with
\be\label{eq:LScalar}
\mathcal{L}\equiv \lambda\ \vartheta^2+\beta \ \nabla_\mu U^\nu\nabla_\nu U^\mu+\alpha \ \dot{U}_\mu\dot{U}^\mu.
\ee
Here $g$ is the metric determinant, $R$ is the Ricci scalar, $S^\mathrm{(mat)}$ is the matter action,
$\dot{U}_\mu\equiv U^\nu\nabla_\nu U_\mu$ and $\vartheta\equiv\nabla_\mu U^\mu$ with $\nabla_\mu$ representing the covariant derivative. The quantities $\alpha$, $\beta$, and  $\lambda$ are coupling constants for khronometric gravity, and we can recover GR by taking the limit $(\alpha, \beta, \lambda) \to 0.$ 
In the action, $G_{\rm bare}$ is the bare gravitational constant, which satisfies
\be 
G_{\rm bare}=G_N\left(1-\frac \alpha 2\right)=\left(1-\frac \alpha 2\right), 
\ee 
where $G_N$ is the Newtonian gravitational constant, measured locally in the solar system~\cite{Carroll:2004ai,Jacobson:2007veq}. We shall set $G_N=1$ throughout this work.
The quantity $U_\mu$ is a fundamental vector of the theory (that would correspond to the \ae ther vector in Einstein-\AE ther theory). This is defined in terms of a ``khronon" scalar $T$ whose constant hypersurfaces define time foliations of spacetime,
and is found to be
\be\label{eq:Ukhronon}
U_\mu=\frac{\nabla_\mu T}{\sqrt{\nabla_\nu T\nabla^\nu T}}.
\ee
$U_\mu$ is also subject to the normalization condition $U^\mu U_\mu=1.$
Thus, we find $U^\mu$ is a unit timelike vector specifying a preferred time direction that classifies khronometric gravity as one that violates Lorentz symmetry. From the above equation, we see that khronometric gravity could be considered to be either a scalar-tensor theory or a vector-tensor theory. This is a result of Forbenius' theorem being bijective. On the one hand, one can start with a scalar that foliates spacetime with its constant hypersurfaces and from this define a hypersurface orthogonal vector field. One the other hand, one can instead start with a vorticity-free vector field and then find a scalar field that foliates spacetime with its constant hypersurfaces~\cite{Poisson:2009pwt}. We give a more detailed analysis of this dual interpretation in Appendix~\ref{app:scalarTensorAnalysis}.

We next present the equations of motion for khronometric gravity. The modified Einstein field equations in this theory, found by varying the action with respect to the metric, are given by~\cite{Barausse:2019yuk} 
\ba\label{eq:EinsteinEqnsTensor}
E_{\mu \nu}\equiv 
G_{\mu\nu}-8\pi G_{\rm bare} T^\mathrm{(mat)}_{\mu\nu}-T^{(k)}_{\mu\nu}=0.
\ea
Here, $G_{\mu\nu} = R_{\mu\nu} - R\, g_{\mu\nu}/2$ is the Einstein tensor with $R_{\mu\nu}$ representing the Ricci tensor and $T^\mathrm{(mat)}_{\mu\nu}$ is the matter stress-energy tensor while $T^{(k)}_{\mu\nu}$ is the stress-energy tensor for the vector field $U^\mu$ defined by
\ba\label{eq:AEStressEnergyTensor}
T^{(k)}_{\mu\nu}=\nabla_\rho\left[J_{(\mu}{}^\rho U_{\nu)}-J^\rho{}_{(\mu} U_{\nu)}-J_{(\mu\nu)} U^{\rho} \right]+\alpha\ \dot{U}_\mu\dot{U}_\nu\nonumber\\
+\left(U_\rho\nabla_\sigma J^{\sigma\rho}-\alpha\ \dot{U}_\rho\dot{U}^\rho\right)U_\mu U_\nu+\frac{\mathcal L}{2}g_{\mu\nu}+2\text{\AE}_{(\mu}U_{\nu)},\nonumber\\
\ea
with
\ba
\label{eq:JTensor}
J^{\mu}{}_{\nu}&
\equiv &\lambda\, \vartheta\delta^\mu_\nu+\beta\ \nabla_\nu U^\mu+\alpha\ \dot{U}_\nu U^\mu,\\
\label{eq:AEVector}
\text{\AE}_\mu& \equiv &\left(g_{\mu\nu}-U_\mu U_\nu\right)\left(\nabla_\rho J^{\rho\nu}-\alpha \ \dot{U}_\rho\nabla^\nu U^\rho\right).
\ea
The vector field satisfies an \ae ther equation of motion,
\be\label{eq:AEEOM}
\nabla_\mu\left(\frac{\text{\AE}^\mu}{\sqrt{\nabla^\nu T \nabla_\nu T}}\right)=0,
\ee
which can be obtained by varying the action with respect to the khronon scalar $T$. One can equivalently derive this equation by varying the Einstein-\ae ther theory action with respect to the vector field with five Lagrange multipliers, one to enforce the unit timelike constraint and four to enforce the zero-vorticity condition~\cite{Yagi:2013ava}. 

Let us now review existing bounds on the theory~\cite{Gumrukcuoglu:2017ijh}. The propagation speed of tensor modes is given by $1/(1-\beta)$, whose deviation from the speed of light has been constrained to be $\sim 10^{-15}$ from the gravitational-wave observation of GW170817 and its electromagnetic counterpart~\cite{LIGOScientific:2017vwq,LIGOScientific:2017zic}. This leads to the constraint $|\beta| \lesssim 10^{-15}$. From other observations and theoretical requirements, such as the stability of the theory, solar system experiments,  Big Bang nucleosynthesis, and cosmological constraints~\cite{Gumrukcuoglu:2017ijh,Afshordi:2009tt}, the remaining coupling constants are constrained to be $|\alpha| \lesssim 10^{-7}$ and $|\lambda| \lesssim 0.01$. Therefore, we can see that $\lambda$ remains relatively unconstrained while the magnitude of $\alpha$ and $\beta$ are more stringently bounded. The main interest of this paper is to study the I-Love-Q relations in khronometric gravity in the $\abzero$ limit, while keeping $\lambda$ free. This is in fact the only parameter choice that makes black holes nonpathological at the universal horizon for all propagation modes~\cite{Ramos:2018oku}. Additionally, we note that the order in which one takes $\alpha$ and $\beta$ to zero does not alter the results throughout this work.

\section{Metric, Vector, and Matter Perturbations}\label{sec:metric_vector_matter_perturbations}
In the following sections, we consider neutron star perturbations due to slow rotation and tidal deformation. We first present the general metric and vector field ansatz, which can be reduced to the slow rotation case or the tidal deformation case by keeping the relevant free functions and spherical harmonic modes for each case. We next show the matter contribution.
\subsection{Metric and Vector Ansatz}
Using the generic, static and spherically symmetric metric as a background with additional terms from appropriate parity perturbations in the Regge-Wheeler gauge~\cite{Regge:1957td,Thorne:1967}, we start with the following ansatz~\cite{Yagi:2013awa} that includes $l=1$ odd and $l=2$ even perturbations: 
\ba\label{eq:generalMetricBeforeCT}
ds^2&=&e^{\nu(r)}[1+\varepsilon^2\ \kappa H_0(r) Y_{2m}(\theta,\phi)]d\tilde t^2\nonumber\\ &-&
e^{\mu(r)}[1-\varepsilon^2\ \kappa H_2(r) Y_{2m}(\theta,\phi)]dr^2\nonumber\\ &-&
r^2[1-\varepsilon^2\ \kappa K(r)Y_{2m}(\theta,\phi)]\nonumber\\ &\times&\{d\theta^2+\sin^2\theta [d\phi-\varepsilon[\Omega_\star-\omega(r)P_1'(\cos\theta)]d\tilde t]^2\}\nonumber\\ &&
+2\varepsilon^2[\kappa \tilde H_1(r) Y_{2m}(\theta,\phi)]d\tilde tdr+\mathcal{O}(\varepsilon^3).
\ea
Here, $\varepsilon$ is a bookkeeping parameter denoting the order of the perturbation, $Y_{2m}(\theta,\phi)$ is the $l=2$ spherical harmonic function, $P_1'(\cos\theta)={dP_1(\cos\theta)}/{d(\cos\theta)}$, $\Omega_\star$ is the (constant)  angular velocity, and 
$\kappa=2\sqrt{\pi/5}$ (chosen so that $\kappa \ Y_{20}(\theta,\phi)=P_2(\cos\theta)$).  
For vector perturbations, we follow
Eq.~(13) in~\cite{Pani:2013wsa} to form our ansatz, 
\ba
\label{eq:generalAEtherBeforeCT}
U_\mu d\tilde x^\mu&=&e^{\nu/2}\bigg\{\left[1+\varepsilon^2\ \kappa X(r) Y_{2m}(\theta,\phi)\right]d\tilde t\nonumber\\ &+&
\varepsilon^2\ \kappa W(r) Y_{2m}(\theta,\phi)dr
+\varepsilon^2\ \kappa V(r) \partial_\theta Y_{2m}(\theta,\phi)d\theta\nonumber\\ &+&
\left[\varepsilon\ S(r)\sin^2\theta+\varepsilon^2\ \kappa V(r) \partial_\phi Y_{2m}(\theta,\phi)\right]d\phi\bigg\}\nonumber\\ &+&\mathcal{O}(\varepsilon^3).
\ea
By normalizing the vector's magnitude to be unity, we find $X=H_0/2.$
We only consider perturbative terms up to quadratic order in $\varepsilon$ in this paper. 
The above ansatz may also be used for tidal perturbations by only considering even parity perturbations; there is no tidal perturbation at $\mathcal{O}(\varepsilon)$ and the leading perturbation enters at $\mathcal{O}(\varepsilon^2)$.

We can further modify this general framework by using the zero-vorticity condition of khronometric gravity to find an appropriate coordinate transformation that simplifies the ansatz~\cite{Yagi:2013ava}. In khronometric gravity, there is a required hypersurface orthogonality condition which can be expressed as requiring  the vorticity vector, $w^\mu,$ to vanish~\cite{Yagi:2013ava}:
\be \label{eq:vorticityDef}
w^\mu \equiv \varepsilon^{\mu\nu\rho\sigma}U_\nu\partial_\rho U_\sigma=0\,,
\ee 
where $\varepsilon^{\mu\nu\rho\sigma}$ is the Levi-Civita tensor.
Using this condition, we can find a coordinate transformation such that there is only one nonvanishing component in the vector ansatz. 
For ansatz shown in Eqs.~\eqref{eq:generalMetricBeforeCT} and~\eqref{eq:generalAEtherBeforeCT}, we find the nontrivial vorticity vector components to be 
\allowdisplaybreaks
\ba\label{eq:generalVorticity}
w^r&=& \varepsilon\ \frac{2e^{\frac{\nu-\mu}{2}}}{r^2}S\cos\theta+\mathcal{O}(\varepsilon^3),\\
w^\theta&=&- \varepsilon\frac{\ e^{\frac{\nu-\mu}{2}}}{r^2\sin\theta}\left[\sin^2\theta\partial_r S\right. \nonumber \\ &&\left.+\varepsilon\ m\sin (m\phi)P^m_2(\cos\theta)(W-\partial_rV)\right]\nonumber\\ &+&\mathcal{O}(\varepsilon^3),\\
w^\phi&=& \varepsilon^2\frac{e^{\frac{\nu-\mu}{2}}}{r^2\sin\theta}\cos (m\phi)P^m_2{}'(\cos\theta)(W-\partial_r V)+\mathcal{O}(\varepsilon^3), \nonumber \\
\ea
where $P^m_2(\cos{\theta})$ is the $l=2$ associated Legendre polynomial and  $P^m_2{}'(\cos\theta)={d P^m_2(\cos\theta)}/{d(\cos\theta)}.$ We can then conclude that $S=0$ and $W(r)=\partial_r V(r)$ in order to eliminate the nonvanishing vorticity terms. 

This allows us to rewrite the vector field as
\ba
\label{eq:generalAEtherVorticityFree}
U_\mu d\tilde x^\mu&=&e^{\nu/2}\bigg\{\left[1+\varepsilon^2\ \kappa \frac{H_0}{2}(r) Y_{2m}(\theta,\phi)\right]d\tilde t\nonumber\\ &+&
\varepsilon^2\ \kappa \partial_r V(r) Y_{2m}(\theta,\phi)dr
+\varepsilon^2\ \kappa V(r) \partial_\theta Y_{2m}(\theta,\phi)d\theta\nonumber\\ &+&
\varepsilon^2\ \kappa V(r) \partial_\phi Y_{2m}(\theta,\phi)d\phi\bigg\}+\mathcal{O}(\varepsilon^3).
\ea
Now, we choose the coordinate transformation given by 
\be \label{eq:coordinateTransformation}
t=\tilde{t}+\varepsilon^2\ 
\kappa V(r)Y_{2m}(\theta,\phi),
\ee 
where, upon differentiating both sides, solving for $d\tilde t$, and substituting the expression into Eq.~\eqref{eq:generalAEtherVorticityFree}, all but the $t$ component of the vector field is eliminated. \sid{We note that this transformed time coordinate corresponds to the perturbed khronon scalar, which is discussed in Appendix~\ref{app:scalarTensorAnalysis}.} 

We now present the metric and vector forms used throughout this paper. After performing the coordinate transformation above, the metric and \ae ther vector ansatz are given by  
\bw
\ba\label{eq:generalMetric}
ds^2&=&e^{\nu(r)}[1+\varepsilon^2\ \kappa H_0(r) Y_{2m}(\theta,\phi)]dt^2-
e^{\mu(r)}[1-\varepsilon^2\ \kappa  H_2(r) Y_{2m}(\theta,\phi)]dr^2\nonumber\\
&-&r^2[1-\varepsilon^2\ \kappa K(r)Y_{2m}(\theta,\phi)]\{d\theta^2+\sin^2\theta [d\phi-\varepsilon[\Omega_\star-\omega(r)P_1'(\cos\theta)]dt]^2\}\nonumber\\ &+& 2\varepsilon^2\ \kappa \left\{ H_1(r) Y_{2m}(\theta,\phi) dtdr - e^{\nu(r)} V(r) [\partial_\theta Y_{2m}(\theta,\phi)dtd\theta + \partial_\phi Y_{2m}(\theta,\phi)dtd\phi]\right\}+\mathcal{O}(\varepsilon^3),
\ea
\ew
\be\label{eq:generalAEther}  
U_\mu dx^\mu=e^{\frac{\nu(r)}{2}}\left[1+\varepsilon^2\ \kappa \frac{H_0(r)}{2} Y_{2m}(\theta,\phi)\right]dt+\mathcal{O}(\varepsilon^3),
\ee
where we define 
\be \label{eq:H1redef}
H_1\equiv\tilde H_1-e^\nu \partial_r V(r).
\ee
Notice that the vector field only has a $t$ component while there are additional $(t,r)$ and $(t,\theta)$ components in the metric.

We now summarize which field functions and spherical harmonic modes are considered at each order of perturbation in the slow rotation or tidal deformation case. We can see that the only radial functions at $\mathcal{O}(\varepsilon^0)$ are $\nu$ and $\mu$, while at $\mathcal{O}(\varepsilon)$ we have $\omega$, and at $\mathcal{O}(\varepsilon^2)$ we have $H_0$, $H_1$, $H_2$, $K$, and $V$. For metric perturbations due to slow rotation, we only have $(l,m)=(1,0)$ modes at $\mathcal{O}(\varepsilon)$ and both $(l,m)=(2,0)$ and $(l,m)=(0,0)$ modes at $\mathcal{O}(\varepsilon^2)$. Tidal deformability only enters at $\mathcal{O}(\varepsilon^2)$ with $l=2$ and all $m$ modes, so when solving the tidal field equations, we neglect $\mathcal{O}(\varepsilon)$ terms in Eq.~\eqref{eq:generalMetric}~\cite{Yagi:2013awa}. Thus, the free field functions are \{$\nu$, $\mu$, $\omega$, $H_0$, $H_1$, $H_2$, $K$, $V$\} in the slow rotation case and \{$\nu$, $\mu$, $H_0$, $H_1$, $H_2$, $K$, $V$\} in the tidal deformation case.

\subsection{Matter Contribution}
In this subsection, we develop the form of the matter stress-energy tensor. First, we discuss the four-velocity of the neutron star. We assume the neutron star is a perfect fluid, with its four-velocity given by
\be\label{eq:4velocity}
u^\mu \partial_\mu=u^{\tilde{t}}(\partial_{\tilde{t}}+\varepsilon\ \Omega_\star\partial_\phi),
\ee
where $\Omega_\star$ is the neutron star angular velocity. Using the normalization condition $u^\mu u_\mu=1$, we find
\be
u^{\tilde{t}}=e^{\frac{\nu}{2}}+\frac{\varepsilon^2}{2}e^{-\frac{3\nu}{2}}\left[(\omega r\sin\theta)^2-\kappa \ e^\nu H_0Y_{2m}\right]+\mathcal{O}(\varepsilon^3).
\ee

\sid{Before proceeding, let us make two observations. First, note that one does not need to perturb the fluid four-velocity more generally because of the same arguments used in GR to construct rotating stars~\cite{Friedman:2013xza}. We have here assumed that the matter fields are barotropic, since neutron stars are expected to be ``cold'' (i.e. the star's temperature is much lower than the Fermi temperature). We can also use the fact that the timelike killing vector $\xi^\alpha=(\partial_t)^\alpha$ and the axisymmetric killing vector $\psi^\alpha=(\partial_\phi)^\alpha$ are both also symmetries of the matter fields, given explicitly by the conditions 
\ba 
\pounds_{\xi}u^\alpha&=&0; \quad \pounds_{\xi}\epsilon=\pounds_{\xi}p=0,\\
\pounds_{\psi}u^\alpha&=&0; \quad \pounds_{\psi}\epsilon=\pounds_{\psi}p=0.
\ea
Here, $\epsilon$ is the energy density, $p$ is the pressure, and $\pounds_v$ denotes the Lie derivative along a vector $v^\alpha$.  Barotropic stars with these symmetries can be modeled with a circular four-velocity, which is precisely the form in Eq.~\eqref{eq:4velocity}~\cite{Friedman:2013xza}. The inclusion of a stationary and axisymmetric \ae ther vector field would not change any of the above, so we maintain the same four-velocity as in GR.}

\sid{Second, note that the coordinate transformation given in Eq.~\eqref{eq:coordinateTransformation} does not change the form of the fluid four-velocity, as can be shown by the vector transformation law. Let tilde indices denote the coordinates used before the transformation given in Eq.~\eqref{eq:coordinateTransformation}, explicitly given by $x^{\tilde \mu}=\{\tilde{t}=t-\varepsilon^2\kappa V(r)Y_{lm}(\theta,\phi),r,\theta,\phi\}$. Thus, we find that 
\ba 
u^{\mu}\partial_{\mu}&=&\frac{\partial x^{\mu}}{\partial x^{\tilde{\mu}}}u^{\tilde{\mu}}\partial_{\mu}=\left(\frac{\partial x^{\mu}}{\partial \tilde t}u^{\tilde t}+\frac{\partial x^{\mu}}{\partial \tilde \phi}u^{\tilde\phi}\right)\partial_{\mu}\nonumber \\
&=&u^{\tilde t}\left[\delta^{\mu}_{t}+\varepsilon \Omega_\star\left( \delta^{\mu}_{t}\frac{\partial t}{\partial \tilde \phi}+\delta^{\mu}_{\phi}\frac{\partial \phi}{\partial \tilde\phi}\right)\right]\partial_{\mu}\nonumber \\
&=&u^{\tilde t}\left(\partial_{t}+\varepsilon \Omega_\star\partial_{\phi}\right)+\mathcal O(\varepsilon^3),
\ea 
where we have here used the fact that ${\partial t}/{\partial \tilde \phi}=\mathcal O(\varepsilon^2)$}.

With this in mind, the stress-energy tensor up to $\mathcal{O}(\varepsilon^2)$ is given by 
\ba
T^\mathrm{(mat)}_{\mu\nu}&=&[\tilde{\epsilon_0}+\tilde{p_0}+\varepsilon^2\ \kappa (\tilde{\epsilon_2}+\tilde{p_2})Y_{2m}]u_\mu u_\nu\nonumber\\ &-& (\tilde{p_0}+\varepsilon^2\ \kappa\, \tilde{p_2}Y_{2m})g_{\mu\nu}+\mathcal{O}(\varepsilon^3),
\ea
where $\tilde{\epsilon_i}$ and $\tilde{p_i}$ are the $i$th order energy densities and pressures, respectively. 
We rescale these matter fields via~\cite{Yagi:2013ava}
\be 
\epsilon_i\equiv \frac{2-\alpha}{2} \tilde{\epsilon_i},\quad p_i\equiv  \frac{2-\alpha}{2} \tilde{p_i},
\ee 
which absorbs the overall factor introduced by $G_{\rm bare}=1-\alpha/2$ in Eq.~\eqref{eq:action}.

\section{I-Love-Q}\label{sec:i_love_q}
In this section, we prove that the I-Love-Q trio in khronometric gravity are not modified from their GR values when $\abzero$. To do so, we use the ansatz in the previous section and consider slowly-rotating or weakly-tidally-deformed neutron stars. The moment of inertia, quadrupole moment and tidal Love number can be computed from the first order in rotation, second order in rotation and first order in tidal deformation solutions, respectively~\cite{Yagi:2013awa}.
The background, modified Tolman-Oppenheimer-Volkoff (TOV) equations with no rotation and tidal deformation are given in Eqs.~\eqref{eq:dMdr}--\eqref{eq:dpdr}, and they only depend on $\alpha$. Thus, the TOV equations are identical to the GR ones in the $\alpha \to 0$ limit. We study below slow rotation and small tidal deformation in turn in the limit of $\abzero$. The full field equations without this restriction can be found in Appendix~\ref{sec:AppendixUncontrainedValues}.

\subsection{Moment of Inertia}\label{subsect:1RotFieldEqns}

We begin by studying how $\lambda$ enters the field equations at first order in spin.
From the action in Eq.~\eqref{eq:action}, we see that $\lambda$ always enters multiplied by $\vartheta\equiv\nabla_\mu U^\mu$. However,
\begin{align}
\label{eq:div_U}
\vartheta=& \nabla_\mu U^\mu= \frac{\partial_\mu (\sqrt{-g} U^\mu)}{\sqrt{-g}} \nonumber \\
= &\frac{1}{\sqrt{-g}}[\partial_t (\sqrt{-g} U^t)+\partial_\phi (\sqrt{-g} U^\phi) ] +\mathcal{O}(\epsilon^2)\nonumber \\
=& \mathcal{O}(\epsilon^2)\,.
\end{align}
In the second equality, we used the fact that the only nonvanishing components on $U^\mu$ (not to be confused with $U_\mu$) are $t$ and $\phi$ at $\mathcal{O}(\varepsilon)$, while the third equality holds from the vector field being stationary and axisymmetric. Because the divergence of $U^\mu$ is the expansion (i.e.~ the rate of change of the cross-sectional area of vector field congruences $U^\mu$)~\cite{Jacobson:2013xta}, physically, the $\lambda$ dependence is absent because the expansion of $U^\mu$ is zero for a rotating configuration, at least to first order in spin. Since the field equations at $\mathcal{O}(\epsilon)$ do not depend on $\lambda$, the moment of inertia is the same as in GR in the $\abzero$ limit. One can also check this finding by looking at the field equations explicitly. The only nonvanishing component of the field equations is given in Eq.~\eqref{eq:spin1} which only depends on $\alpha$ and $\beta$. Therefore, it reduces to the GR one when these constants are set to zero. 

\subsection{Quadrupole Moment}
\label{subsect:2RotFieldEqns}

We next consider the field equations at second order in spin to find the $\lambda$ dependence on the quadrupole moment.
In this subsection, we use the metric exactly as shown in Eq.~\eqref{eq:generalMetric}. Note that since $g^{tr}$ and $g^{t\theta}$ are nonvanishing, the contravariant form of the vector $U^\mu$ acquires an $r$ and $\theta$ dependence in the $U^r$ and $U^\theta$ components, respectively. This means that the divergence of the vector is nonzero and thus $\lambda$ enters into the vector field equations. 
The modified field equations now depend on the following set of functions: $\{H_0,H_1,H_2,V,K,p_2,\epsilon_2\}$, which depend only on $r$.

Let us examine the modified field equations analytically in order to discuss the nontrivial appearances of $\lambda$. 
When considering the $\abzero$ limit in Eq.~\eqref{eq:AEStressEnergyTensor}, we obtain
\ba
T^{(k)}_{\mu\nu}&=&\nabla_\rho\left[J_{(\mu}{}^\rho U_{\nu)}-J^\rho{}_{(\mu} U_{\nu)}-J_{(\mu\nu)} U^{\rho} \right]+\frac{\mathcal L}{2}g_{\mu\nu}\nonumber\\
&+&U_\rho\nabla_\sigma J^{\sigma\rho}U_\mu U_\nu+2\text{\AE}_{(\mu}U_{\nu)},
\ea
where 
\ba J^\mu {}_\nu&=&\lambda\ \vartheta\delta^\mu_\nu,\\
 \mathcal L&=&\lambda\ \vartheta^2,\\
 \vartheta&=&\nabla_\rho U^\rho=\nabla_r U^r+\nabla_\theta U^\theta.
 \ea
 Since $J^\mu {}_\nu=J_\nu{}^\mu,$ we can see clearly 
\be 
J_{(\mu}{}^\rho U_{\nu)}-J^\rho{}_{(\mu} U_{\nu)}=0.
\ee
Next, we note that $J^{\mu\nu}=\lambda\ \vartheta\ g^{\mu\nu},$ which gives $\text{\AE}_\mu=\left(g_{\mu\nu}-U_\mu U_\nu\right)\nabla_\rho J^{\rho\nu}.$ Using these equations, we find 
\ba 
\label{eq:T^k}
T^{(k)}_{\mu\nu}&=&\lambda\left[( U_\mu U_\nu-  g_{\mu\nu})\pounds_U\vartheta+\frac{\vartheta^2}{2}g_{\mu\nu}\right]\nonumber+2\text{\AE}_{(\mu}U_{\nu)}, \\\\
\label{eq:AE_mu}
\text{\AE}_\mu&=&\lambda\ (\nabla_\mu \vartheta-U_\mu \pounds_U\vartheta\, ).
\ea
Here, $\pounds_U$ denotes a Lie derivative with respect to the vector field. 

Let us now analyze each term of $T^{(k)}_{\mu\nu}$ to reveal the $\lambda$ dependence. Taking a closer look at the Lie derivative term, we find
\be
\pounds_U\vartheta=U^\mu\nabla_\mu\vartheta=U^r\nabla_r\vartheta+U^\theta\nabla_\theta\vartheta=\mathcal O(\varepsilon^4),
\ee
where we used the fact that $\vartheta$ is independent of $t$ and $\phi$ and that $U^r$, $U^\theta$, and $\vartheta$ are all $\mathcal{O}(\varepsilon^2).$
Therefore, we can neglect all terms containing the Lie derivative of $\vartheta$ with respect to the vector field. Similarly, we see the last term in the brackets of Eq.~\eqref{eq:T^k} is proportional to $\vartheta^2=\mathcal{O}(\varepsilon^4)$. Thus, at this order of perturbation, we are left with 
\be
T^{(k)}_{\mu\nu}=2 \lambda \nabla_{(\mu} \vartheta \, U_{\nu)}\,.
\ee
Again using the static and axisymmetric property of $\vartheta$, the only nonvanishing components of $\nabla_{\mu} \vartheta$ are $\mu=(r, \theta)$. Keeping in mind that $U_\mu$ only has a nonvanishing $t$ component, we only find $\lambda$ terms in the $(t,r)$ and $(t,\theta)$ terms of $T^{(k)}_{\mu\nu}$. This means that only $E_{tr}=0$ and $E_{t\theta}=0$ contain $\lambda$ in the modified field equations.

Let us now discuss how the quadrupole moment depends on $\lambda$. Such a moment can be derived by looking at the asymptotic behavior of $g_{tt}$ at large $r$, which is determined through $H_0$ and $K$. As we discuss in more detail in Appendix~\ref{app:2nd}, it turns out that the diagonal components of $E_{\mu\nu}=0$ and $E_{r\theta}=0$ give coupled equations for $H_0$, $K$ and $H_2$. Based on the discussion earlier in this subsection, these components do not depend on $\lambda$. Hence, the quadrupole moment reduces to that of GR in the $\abzero$ limit. We have confirmed this result by directly deriving equations for $H_0$, $K$ and $H_2$, which are given in Eqs.~\eqref{eq:H2_rot}--\eqref{eq:dH0dr_rot}. These equations only depend on $\alpha$ and $\beta$, and thus reduce to the GR field equations when we set these coupling constants to zero.

\subsection{Tidal Love Number}\label{sec:tidal}

Finally, we study tidal perturbations and how the tidal Love number depends on $\lambda$. We focus on the leading, even-parity perturbations. This amounts to neglecting terms that are $\mathcal{O}(\varepsilon)$ and keeping those at $\mathcal{O}(\varepsilon^2)$ in Eq.~\eqref{eq:generalMetric}. In practice, one can achieve this by taking the field equations at second order in spin and setting $\omega = 0$
Doing so, we find that the discussion in Sec.~\ref{subsect:2RotFieldEqns} still holds for tidal perturbations and, in particular, $\lambda$ does not enter the diagonal components of $E_{\mu\nu}$ or $E_{r\theta}$.

The tidal Love number is defined as the ratio between the tidally-induced quadrupole moment and the external tidal field. Similar to extracting the (rotation-induced) quadrupole moment described in Sec.~\ref{subsect:2RotFieldEqns}, both of these quantities can be read off by looking at the asymptotic behavior of $g_{tt}$ at large $r$. The only difference is that one does not impose asymptotic flatness to extract the tidal field strength, since the expansion is formally done in an intermediate buffer zone (i.e.~in a spatial region that is not too close to the neutron star surface but also not too close to the object causing the tidal deformation). Given that $H_0$ and $K$ are independent of $\lambda$, the tidal Love number is the same as the GR one in the limit $\abzero$. Once again, we checked this result explicitly by looking at the field equations for $H_0$ directly. For tidal perturbations, we find that $H_0 = H_2 \equiv H$ due to the absence of the $\mathcal{O}(\varepsilon)$ perturbation; the equation for $H$ is given in Eq.~\eqref{eq:H1_generic}, which only depends on $\alpha$ (unlike the field equations at second order in spin which also depend on $\beta$). This equation, therefore, reduces to the GR one in the limit $\alpha \to 0$.

\section{Electric-type Perturbation to the Shift}
\label{sec:V_H1_analysis}

Are slowly-rotating neutron stars or tidally-perturbed neutron stars in khronometric gravity the same as those in GR when $\abzero$? The answer to this question is not trivial as there are some components in the modified Einstein equations (not relevant for extracting I-Love-Q) that have $\lambda$ dependence and that do not explicitly vanish in the $\abzero$ limit.
Below, we study slowly-rotating stars and tidally-deformed ones in turn.

\subsection{Slow rotation}\label{subsec:analyticSlowRot}

We begin by focusing on slowly-rotating neutron stars to second order in spin and focus on solving the $\lambda$-dependent part of the equations \emph{analytically} within the PM approximation. 

\subsubsection{Field Equations}\label{subsec:V_H1_analysis_fieldeq}

We first derive the equations for metric perturbations that depend on $\lambda$.
As we showed in Sec.~\ref{subsect:2RotFieldEqns}, it is the $(t,r)$ and $(t,\theta)$ components of the modified Einstein equations that contain the $\lambda$ dependence. These two equations give a coupled system for $H_1$ and $V$. When taking the $\abzero$ limit (see Eqs.~\eqref{eq:H1_generic} and~\eqref{eq:V_generic} for the full equations), we find
\bw
\ba\label{eq:d2H1dr2}
 \frac{d^2H_1}{dr^2} &=&\frac{2}{r(r-2M)}(2\pi r^3 p_0+6\pi r^3 \epsilon_0+M-r)  \frac{dH_1}{dr}-\frac{3e^\nu(2\lambda+1)}{\lambda r(r-2M)} \frac{dV}{dr}-\frac{6e^\nu(4\pi r^3 p_0+5M-2r)}{r^2(r-2M)^2}V
+\left\{\frac{4\pi r^2}{r-2M}\frac{d\epsilon_0}{dr}\right.\nonumber\\&+&\left.\frac{9\lambda M^2+2r(6\pi \lambda r^2p_0+18\pi \lambda r^2 \epsilon_0 +4\lambda-3)M+r^2[8\pi \lambda r^2(2\pi \epsilon_0 r^2-1)p_0-16\pi r^2 \lambda \epsilon_0-2\lambda+3]}{\lambda r^2 (r-2M)^2}\right\}H_1,
\\
\label{eq:d2Vdr2}
\frac{d^2V}{dr^2} &=& \frac{4\pi r^3(\epsilon_0-3 p_0)-4M}{r(r-2M)}\frac{dV}{dr}-\frac{12\lambda}{r(r-2M)} V+2e^{-\nu}\frac{(3\lambda-1)M+2\pi r^3[(2\lambda+1)\epsilon_0- p_0]-2\lambda r}{r(r-2M)}H_1-(2\lambda+1)\frac{dH_1}{dr}.\nonumber\\
\ea
\ew
Here, $M(r)$ is defined as
\be
M(r)\equiv\frac{r}{2}\left[1-e^{-\mu(r)}\right].
\ee

Before solving the above equations, let us discuss taking the GR limit.
When taking the limit $\lambda \to 0$, Eq.~\eqref{eq:d2H1dr2} has an apparent divergence. However, by multiplying both sides  by $\lambda$, we find the condition
\be\label{eq:H1Lim}
H_1=-e^\nu\frac{dV}{dr}+\mathcal{O}(\lambda),
\ee
when $\lambda \ll 1$.
This condition identically satisfies Eq.~\eqref{eq:d2Vdr2} in the limit 
$\lambda \to 0$, making it vanish at $\mathcal O(\lambda^0).$ This indicates that the system loses a degree of freedom in the GR limit, since the pair of coupled differential equations collapses into one condition that leaves $H_1$ and $V$ undetermined. However, in GR there is no vector field to give a condition on $V$, which originated as a vector perturbation in Eq.~\eqref{eq:generalAEtherBeforeCT}. Therefore, the GR limit is actually $\tilde H_1=0$ in Eq.~\eqref{eq:generalMetricBeforeCT}, which describes the metric before the time coordinate transformation in Eq.~\eqref{eq:coordinateTransformation}. Thus, $V$ in the metric ansatz in Eq.~\eqref{eq:generalMetric} is simply an artefact of the coordinate transformation we performed in Eq.~\eqref{eq:coordinateTransformation}. By comparing Eqs.~\eqref{eq:H1redef} and~\eqref{eq:H1Lim}, we see that $\tilde H_1=0,$ and transforming back to the original time coordinate removes all terms dependent on $V$. Thus, we recover the correct GR limit, and we find that $V$ becomes purely a gauge artefact from the transformation given by Eq.~\eqref{eq:coordinateTransformation}.   

\subsubsection{PM Expansion}\label{subsect:PMExpansion}

We wish to solve Eqs.~\eqref{eq:d2H1dr2} and~\eqref{eq:d2Vdr2} both in the interior and exterior of neutron stars under the boundary conditions of regularity at the center and asymptotic flatness at infinity. Since the system is composed of two second-order differential equations, there are four integration constants for each region. Two of them are associated with terms that grow in $r$, while the other two correspond to solutions that decay in $r$. Imposing asymptotic flatness at infinity eliminates the first two modes in the exterior, while the regularity at the center removes the other two in the interior. Given that Eqs.~\eqref{eq:d2H1dr2} and~\eqref{eq:d2Vdr2} are homogeneous, there is no other contribution to the solution, and upon matching the solutions at the surface, we find $H_1 = V = 0$.

To show the above argument explicitly, we consider solving Eqs.~\eqref{eq:d2H1dr2} and~\eqref{eq:d2Vdr2} analytically within the PM approximation~\cite{Gupta:2021vdj,Yagi:2021loe}.
We expand the TOV equations along with Eqs.~\eqref{eq:d2H1dr2} and~\eqref{eq:d2Vdr2} in orders of compactness $\mathcal{C}=M_\star/R_\star$, where $M_\star$ and $R_\star$ are the stellar mass and radius, respectively. The leading contribution of each background function is as follows: $M = \mathcal{O}(\mathcal{C}')$, $\nu = \mathcal{O}(\mathcal{C}')$ and $p_0=\mathcal{O}(\mathcal{C}'^2)$~\cite{Gupta:2021vdj,Yagi:2021loe}. Here $\mathcal{C}'$ is a  bookkeeping parameter which denotes the order of a term in compactness. For $H_1$ and $V$, we use the following ansatz:
\begin{equation}
H_1(r)=\sum_{j=0} \eta_j(r) \ \mathcal{C}'^j, \quad V(r)=\sum_{j=0} v_j(r) \ \mathcal{C}'^j.
\end{equation}
We substitute the above ansatz in Eqs.~\eqref{eq:d2H1dr2} and~\eqref{eq:d2Vdr2}
and expand in powers of $\mathcal{C}'$. We solve the equations for $\eta_j$ and $v_j$ order by order, both in the interior and exterior regions. We impose regularity at the center and infinity, together with the following boundary conditions at the surface for the continuity and smoothness of the solutions:
\ba\label{eq:PMBC1}
\eta_j^{\text{(int)}}(R_\star)= \eta_j^{\text{(ext)}}(R_\star),\\
\label{eq:PMBC3}
\eta'_j{}^{\text{(int)}}(R_\star)=\eta'_j{}^{\text{(ext)}}(R_\star), \\
\label{eq:PMBC4}
v_j^{\text{(int)}}(R_\star)=v_j^{\text{(ext)}}(R_\star),\\
\label{eq:PMBC2}
v'_j{}^{\text{(int)}}(R_\star)=v'_j{}^{\text{(ext)}}(R_\star),
\ea
where the primes indicate a derivative with respect to $r$.

We now show that $H_1$ and $V$ vanish through the method of induction. We take the following steps: (i) show that $H_1 = V = 0$ when $j=0$; (ii) assume that $H_1 = V=0$ for $j \leq n - 1$ and show that $H_1 = V = 0$ holds for $j=n$.

Let us start by looking at the first step. When $j=0$, the equation for $\eta_0$ and $v_0$ are given by 
\ba
\label{eq:1orderPMeta}
\eta''_0&=&\frac{12\lambda v_0-r\left[(3-2\lambda)\eta_0+(3+6\lambda)v'_0+2\lambda r\eta'_0\right]}{\lambda r^3},\nonumber\\
\\
\label{eq:1orderPMv}
v''_0&=&-\frac{12\lambda v_0+4r\lambda \eta_0+r^2(1+2\lambda)\eta'_0}{r^2},
\ea
where we assume $\lambda \neq 0$. 

For the interior, by requiring regularity at the center we find
\ba\label{eq:eta1Int}
\eta^{\text{(int)}}_0&=&-2A_0 r+B_0 r^3,\\
v^{\text{(int)}}_0&=&A_0r^2-\frac{(3+10\lambda)B_0}{12(1+\lambda)} r^3,
\ea
while for the exterior, by requiring regular solutions in the limit $r\to\infty$, we obtain
\ba
\eta^{\text{(ext)}}_0&=&\frac{C_0}{r^2}+\frac{3D_0}{r^4},\\
v^{\text{(ext)}}_0&=&\frac{C_0}{r(1+6\lambda)}+\frac{D_0}{r^3}. \label{eq:v1Ext}
\ea
Here $A_0$, $B_0$, $C_0$ and $D_0$ are integration constants. 
Imposing continuity and differentiability of $\eta_0$ and $v_0$ at the surface yields $A_0=B_0=C_0=D_0=0,$ showing that $H_1 = V = 0$ at $\mathcal{O}(\mathcal{C}'^0)$. 

Next, we study the second step. We assume $H_1 = V=0$ for $j \leq n - 1$ and consider the equations for $\eta_n$ and $v_n$ at $\mathcal{O}(\mathcal{C}'^n)$. Given that $\eta_n$ and $v_n$ are already of $\mathcal{O}(\mathcal{C}'^n)$, we only need coefficients at $\mathcal{O}(\mathcal{C}'^0)$. This means that the equations for $\eta_n$ and $v_n$ are the same as Eqs.~\eqref{eq:1orderPMeta} and~\eqref{eq:1orderPMv}. Then, we can repeat the process explained in the first step and this leads to $H_1 = V = 0$ at $\mathcal{O}(\mathcal{C}'^n)$. This concludes our proof that $H_1 = V = 0$ to all orders in the PM expansion. Notice that this proof does not depend on the choice of the equation of state.

\subsection{Tidal Deformation} \label{subsect:KLoveNumber}

In this subsection, we study tidally-deformed neutron stars in khronometric gravity in more detail. In particular, we focus on  the non-GR fields $V$ and $H_1$ that depend on $\lambda$ but do not affect the usual tidal Love number, computed from the $g_{tt}$ perturbations discussed above. Unlike the analysis for slowly-rotating stars in Sec.~\ref{subsect:PMExpansion}, we do not impose asymptotic flatness on the metric and vector fields (because the presence of the perturber prevents us from doing so), and this allows us to extract tidal terms that grow with $r$. We will find it more insightful to work with $(V,\tilde H_1)$ rather than $(V,H_1)$, where $V$ originated as a vector perturbation, while $\tilde H_1$ is a metric perturbation before the coordinate transformation in Eq.~\eqref{eq:H1redef}. We consider the following two cases and introduce three new Love numbers: (i) two of them are obtained by imposing that the vector field strength be zero (a shift and a vector Love number), and (ii) one of them is found by letting the metric field strength be zero (another vector Love number). Case (i) is similar to previous work on I-Love-Q relations in scalar-tensor theories~\cite{Pani:2014jra}, where the authors did not impose asymptotic flatness on the metric but required the scalar field to be regular as $r\to\infty$ (see
~\cite{Cardoso:2017cfl,Saffer:2021gak} for similar analyses in different theories of gravity). We then discuss the physical interpretation of these Love numbers briefly.

\subsubsection{Khronometric Love number}
We now detail the PM expansion for tidally-deformed neutron stars in khronometric theory. The procedure is similar to the PM analysis in Sec.~\ref{subsect:PMExpansion}, but we now retain growing modes in the exterior to study tidal effects. We note that since Eqs.~\eqref{eq:d2H1dr2} and~\eqref{eq:d2Vdr2} are independent of $\omega$, they apply to the tidal deformation case as well. At zeroth order in compactness, this will yield an interior solution still given by
\ba\label{eq:eta1IntLove}
\eta^{\text{(int)}}_0&=&-2A_0 r+B_0 r^3,\\
v^{\text{(int)}}_0&=&A_0r^2-\frac{(3+10\lambda)B_0}{12(1+\lambda)} r^4,
\ea
but the exterior solution will now take the form
\ba\label{eq:eta1ExtLove}
\eta^{\text{(ext)}}_0&=&\frac{C_0}{r^2}+3\frac{D_0}{r^4}-2F_0r+G_0r^3,\\
\label{eq:v1ExtLove}
v^{\text{(ext)}}_0&=&\frac{C_0}{r(1+6\lambda)}+\frac{D_0}{r^3}+F_0r^2-\frac{(3+10\lambda)G_0r^4}{12(1+\lambda)},\nonumber\\
\ea
where $F_0$ and $G_0$ are additional integration constants characterizing the growing mode.
When matching the solutions at the surface of the star, we will no longer be able to show that these fields vanish. 

To acquire a better physical meaning of each term in the above solutions, we go back to the original coordinate system in Eq.~\eqref{eq:generalMetricBeforeCT} before applying the time transformation in Eq.~\eqref{eq:coordinateTransformation}.
This can be accomplished by simply solving Eq.~\eqref{eq:H1redef} for $\tilde H_1$, which is the $(t,r)$ component radial perturbation in the original coordinates.
In the exterior, because $\nu(r)$ is nonzero only at $\mathcal O(\mathcal C')$, this corresponds to 
\be\label{eq:eta0ExtLoveOriginalC}
\tilde \eta_0^{(\text{ext})}=\eta_0^{(\text{ext})}+\partial_r v^{\text{(ext)}}_0+\mathcal O(\mathcal C') =\frac{6 \lambda C_0}{(1+6\lambda) r^2} -\frac{7\lambda G_0 r^3}{3(1+\lambda)},
\ee
\be
\tilde \eta_0^{(\text{int})}=\eta_0^{(\text{int})}+\partial_r v^{\text{(int)}}_0+\mathcal O(\mathcal C') =-\frac{7\lambda B_0 r^3}{3(1+\lambda)},
\ee
As before, $\mathcal C'$ is a bookkeeping parameter which denotes the order of compactness.
These expressions clearly show that this field vanishes in the GR limit ($\lambda\to 0$), as expected. We see that this metric function only has two terms in the exterior, and we can interpret $C_0$ and $G_0$ as originating from the metric, while the remaining contributions in Eq.~\eqref{eq:eta1ExtLove} originate from the coupling of the metric and vector field.

We shall now provide a procedure to eliminate some of the constants of integration, such that we may define a new shift Love number, similar to the scalar Love number defined in scalar-tensor theories~\cite{Pani:2014jra}. As is, the system has 6 constants of integration (2 in the interior $(A_0,B_0)$, 4 in the exterior $(C_0,D_0,F_0,G_0)$) and 4 boundary conditions (continuity of the field functions and their derivatives). Reference~\cite{Pani:2014jra} had the same number of constants and boundary conditions in scalar-tensor theories, and so the authors defined a scalar Love number by imposing (by hand) that the scalar field be regular in the exterior region (namely no tidal term in the scalar field), which eliminates one of the integration constants. If we were to set all of the growing modes of $V$ to zero in the exterior in khronometric gravity, one would need to eliminate two integration constants ($F_0$ and $G_0$), and 
both $V$ and $\tilde H_1$ would vanish since the system would match the slowly rotating case. Instead, it proves worthwhile to consider the cases in which $F_0=0$ and $G_0=0$ separately. The former corresponds to setting the vector field strength to zero while the latter would set the metric field strength to zero. This will confirm that $C_0$ corresponds to the tidal response from the metric field strength $G_0$ (which is already apparent by Eq.~\eqref{eq:eta0ExtLoveOriginalC}) while $D_0$ is the quadrupolar response due to both field strengths. In either case, we will be left with one undetermined constant, which will cancel upon taking ratios to define new khronometric Love numbers. This allows us to define a shift Love number from $\tilde H_1$ and two vector Love numbers from $V$.

We next discuss the tidal strength that is necessary to compute the Love numbers.
At zeroth order, matching the solutions at the boundary yields only one nonvanishing term, which is interpreted as a tidal field strength. Consider first the case $F_0=0$, whereby we consider only the metric field strength. We label this field ${\mathcal E},$ and 
impose that the tidal field strength term of $\tilde H_1$ is fully captured by the $\mathcal O (\mathcal C^0)$ contribution without loss of generality. Specifically, we impose 
\ba
{\mathcal{E}}&=&\sum_j {\mathcal{E}}_j  \mathcal C'^j={\mathcal{E}}_0.
\ea
Thus, the solution at zeroth order for the fields given in Eqs.~\eqref{eq:eta1IntLove}--\eqref{eq:eta0ExtLoveOriginalC} is $G_0=B_0=-{3(1+\lambda){\mathcal{E}}}/{7
}$
with all other constants vanishing. 
This choice of parametrization keeps the $\lambda$ dependence of $\tilde H_1$ explicit and clearly shows that the field will vanish in the GR limit (as expected). 
The $G_0=0$ case is analogous, where we may define 
\be 
\mathcal{V}=\sum_j \mathcal{V}_j  \mathcal C'^j=\mathcal{V}_0,
\ee 
and the only nonzero constant is $F_0=A_0=\mathcal{V}$ in Eqs.~\eqref{eq:eta1IntLove}--\eqref{eq:eta0ExtLoveOriginalC}. Note that in this case we can show that the metric field $\tilde H_1$ vanishes altogether.

At higher orders of compactness, we must now consider matter fields that couple to lower-order metric and vector fields. To keep our analysis analytically tractable, we will use the Tolman VII model~\cite{Tolman:1939jz}, whose energy density profile is given by
\be 
\epsilon(r)=\epsilon_c\left(1-\frac{r^2}{R_\star^2}\right)=\frac{15M_\star}{8\pi R_\star^3}\left(1-\frac{r^2}{R_\star^2}\right),
\ee 
with $\epsilon_c$ representing the central energy density. This profile allows one to solve the TOV equation analytically, and the solution can accurately model realistic neutron solutions in GR~\cite{Lattimer:2000nx,Jiang:2019vmf}, scalar-tensor theories~\cite{Yagi:2021loe}, and Einstein-\AE ther theory~\cite{Gupta:2021vdj}.

Let us now discuss the three different quadrupole moments induced on the metric and vector fields. At higher orders of compactness, the system of differential equations acquires a source term from lower-order fields. Due to this, we find nonvanishing decaying modes, which include the quadrupole moments of the system. We find that the constant of integration $C_j$ at all orders corresponds to the quadrupole response of the metric field due to the metric field strength ${\mathcal{E}}$. This is clear from the zeroth-order expression of Eq.~\eqref{eq:eta0ExtLoveOriginalC}, but it can be made explicit by matching all of the boundary conditions at all orders, yielding $C_j\propto {\mathcal{E}}.$ We thus label this constant $C_j=(1+6\lambda)\mathcal Q^{({\mathcal{E}},\tilde H_1)}_j/6$, where the first superscript indicates the tidal field strength that induces the moment, while the other indicates to which field the multipole belongs. However, $D_j$ contains responses from both the metric and vector field strengths, which can be seen from the leading-order expression given in Eq.~\eqref{eq:v1ExtLove}. When imposing all of the boundary conditions, this decaying mode depends on both $\mathcal V$ and ${\mathcal E}.$ The former corresponds to multipole responses of the vector field due to the vector field strength, labelled as $\mathcal Q^{(\mathcal V,V)}$, while the latter is the multipole response of the vector field due to the metric field, labeled $\mathcal Q^{({\mathcal{E}},V)}$. We thus take $D_j=\mathcal Q^{({\mathcal{E}},V)}_j+\mathcal Q^{( \mathcal V,V)}_j$. We then construct the full multipole moment by taking
\ba 
\mathcal Q^{({\mathcal{E}},\tilde H_1)}&=&\sum_j \mathcal Q^{({\mathcal{E}},\tilde H_1)}_j\mathcal{C}'^j,\\
\mathcal Q^{({\mathcal{E}},V)}&=&\sum_j\mathcal Q^{({\mathcal{E}},V)}_j\mathcal{C}'^j,\\
\mathcal Q^{(\mathcal V,V)}&=&\sum_j\mathcal Q^{(\mathcal V,V)}_j \mathcal{C}'^j.
\ea

Let us now present the quantities $\tilde H_1$ and $V$, from which we find three khronometric Love numbers. The three Love numbers come from the three quadrupole moments which are induced on the shift metric and vector fields discussed above. 
Finding solutions order by order, we find that the fields up to fourth order in compactness are given by the expressions
\bw 
\ba 
\tilde{H}_1&=&\lambda {\mathcal{E}}r^3\left[1+\mathcal{O}\left(\frac{M_\star}{r}\right)\right]+\lambda\frac{\mathcal Q^{({\mathcal{E}},\tilde H_1)}}{r^2}\left[1+\mathcal{O}\left(\frac{M_\star}{r}\right)\right]+\mathcal {O}\left(\frac{M_\star^5}{R_\star^5}\right),
\label{eq:H1_tilde}
\\
\label{eq:Vtidal}
V&=&\frac{(10 \lambda +3) {\mathcal{E}}}{28}r^4\left[1+\mathcal{O}\left(\frac{M_\star}{r}\right)\right]+{\mathcal V}r^2\left[1+\mathcal{O}\left(\frac{M_\star}{r}\right)\right]\nonumber\\
&&+ \frac{\mathcal Q^{({\mathcal{E}},\tilde H_1)}}{6 r}\left[1+\mathcal{O}\left(\frac{M_\star}{r}\right)\right]+\frac{\mathcal Q^{({\mathcal{E}},V)}+\mathcal Q^{\left(\mathcal V,V\right)}}{r^3}\left[1+\mathcal{O}\left(\frac{M_\star}{r}\right)\right]+\mathcal {O}\left(\frac{M_\star^5}{R_\star^5}\right),
\ea 
where
\ba
\mathcal Q^{({\mathcal{E}},\tilde H_1)}&=&\frac{M_\star R^4_\star  {\mathcal E}\mathcal F}{28},\quad \mathcal Q^{({\mathcal{V}},V)}=-\frac{M_\star R^4_\star\mathcal V\mathcal F}{42 },\quad\mathcal{F} = 1-\frac{652}{143}\frac{M_\star}{R_\star}+\frac{1805957}{255255}\frac{M_\star^2}{R_\star^2}-\frac{1545158}{373065}\frac{M_\star^3}{R_\star^3} \\
\mathcal Q^{({\mathcal{E}},V)}&=&-\frac{ M_\star R^6_\star\lambda\mathcal E }{99}\left[1-\frac {6(837\lambda-4063) } {15925\lambda}\frac{M_\star}{R_\star} \right.\nonumber\\
&&
-\frac {552882330 \lambda ^3 + 4382218319\lambda ^2 + 2914707384 \lambda + 18706545 -232792560\lambda(11\lambda+6)\log R_\star} {246901200  \lambda^2  }\frac{M_\star^2}{R_\star^2}\nonumber\\
&&\left.+\frac {223219861326 \lambda^3 + 513253785137 \lambda ^2+ 346830640200 \lambda +7552551699  -93987310032\lambda(11\lambda+6)\log R_\star  } {21863101260 \lambda ^2}\frac{M_\star^3}{R_\star^3}\right].\nonumber\\
\ea
\ew
Above, the $\mathcal O\left(\frac{M_\star}{r}\right)$ terms come from source terms coupling to lower-order metric and vector fields. Notice that these terms can contaminate the multipolar structure of lower-order modes if they are not separated out like we do above. Thus, we shall consider only the leading-order terms because these give the pure multipolar response without source coupling terms.
To make this separation, one should impose only boundary conditions given by Eqs.~\eqref{eq:PMBC1} and~\eqref{eq:PMBC2} to solve for the constants in the interior, while keeping the $\mathcal Q^{(.,.)}_j$ terms undetermined. Then, one can impose the two remaining boundary conditions order by order to find the above values. The above result confirms that the quantity ${\mathcal E}$ is the metric field strength while the $\mathcal V$ term corresponds to a vector field strength at higher orders in compactness as well; in fact, if one were to take ${\mathcal E}\to 0,$ the metric field vanishes altogether. Thus, its appearance in $V$ comes from the coupling between the vector and metric field.

We now define three new Love numbers.
Consider first the case of $\mathcal V=0.$ We propose defining a shift Love number by selecting the mode we identified above as the moment induced by the metric tidal field.
Namely, we can define a shift Love number for $\tilde H_1$ by taking the ratio of the leading-order growing mode coefficient ${\mathcal E}$ and the coefficient of the decaying mode $\mathcal Q^{({\mathcal{E}},\tilde H_1)}$: 
\bw
\ba 
\Lambda_{\rm shift} \equiv \frac{\mathcal Q^{({\mathcal{E}},\tilde H_1)}}{ {\mathcal{E}}} &=& \frac{M_\star R^4_\star \mathcal F}{28} = \frac{M_\star R_\star^4}{28}\left(1-\frac{652}{143}\frac{M_\star}{R_\star}+\frac{1805957}{255255}\frac{M_\star^2}{R_\star^2}-\frac{1545158}{373065}\frac{M_\star^3}{R_\star^3}\right)+\mathcal O\left(\frac{M_\star^5}{R_\star^5}\right).
\label{eq:tidalLove}
\ea 
This quantity also has consistent units when compared to the usual quadrupolar electric-type Love number from $g_{tt}$. Doing the same with the quadrupole moment induced on $V,$ we define a vector Love number
\ba 
\Lambda_{({\mathcal{E}},V)}&\equiv&\frac{28\mathcal Q^{({\mathcal{E}},V)}}{(10\lambda+3){\mathcal{E}}}=-\frac{28 M_\star R^6_\star\lambda }{99(10\lambda+3)}\left[1-\frac {6(837\lambda-4063) } {15925\lambda}\frac{M_\star}{R_\star} \right.\nonumber\\
&&
-\frac {552882330 \lambda ^3 + 4382218319\lambda ^2 + 2914707384 \lambda + 18706545 -232792560\lambda(11\lambda+6)\log R_\star} {246901200  \lambda^2  }\frac{M_\star^2}{R_\star^2}\nonumber\\
&&\left.+\frac {223219861326 \lambda^3 + 513253785137 \lambda ^2+ 346830640200 \lambda +7552551699  -93987310032\lambda(11\lambda+6)\log R_\star  } {21863101260 \lambda ^2}\frac{M_\star^3}{R_\star^3}\right]\nonumber\\&&+\mathcal O\left(\frac{M_\star^5}{R_\star^5}\right),
\ea
though this Love number will differ in units from the usual electric-type Love number.
In the case where ${\mathcal E}=0,$ the metric vanishes (and therefore we do not define the Love number on $H_1$ induced by the vector tidal field), and we may define another vector Love number, given by 
\ba \label{eq:tidalVectorLove}
\Lambda_{(\mathcal{V},V)}\equiv \frac{\mathcal Q^{({\mathcal{V}},V)}}{\mathcal V}&=&-\frac{M_\star R^4_\star\mathcal F}{42 } = -\frac{M_\star R_\star^4}{42}\left(1-\frac{652}{143}\frac{M_\star}{R_\star}+\frac{1805957}{255255}\frac{M_\star^2}{R_\star^2}-\frac{1545158}{373065}\frac{M_\star^3}{R_\star^3}\right)+\mathcal O\left(\frac{M_\star^5}{R_\star^5}\right).
\ea 
\ew
Note $\Lambda_{(\mathcal{V},V)}$ will have consistent units with the usual Love number as well.
To summarize, we have defined three additional Love numbers, one of which characterizes the response of the metric field due to the metric field strength and two of which characterize the response of the vector field due to both the metric and vector field strengths.

We conclude this analysis by discussing the GR limit of the fields and khronometric Love numbers. We find that the solutions have two branches, one corresponding to the GR limit ($\lambda=0$), and another that corresponds to constrained khronometric gravity ($\lambda\neq0$). This happens because, as discussed at the end of Sec.~\ref{subsec:V_H1_analysis_fieldeq}, the system defined by Eqs.~\eqref{eq:d2H1dr2}--\eqref{eq:d2Vdr2} loses a degree of freedom in the GR limit. This is expected since there is no condition in GR to determine $V,$ which originated from a vector perturbation. Thus, we see that the khronometric branch ($\lambda\neq 0$) predicts non-zero shift and vector Love numbers, as well as a $V$ field, while these quantities are not present in the GR ($\lambda=0$) branch. 
Note that the above expression for $V$ in Eq.~\eqref{eq:Vtidal} diverges in the GR limit, but this is not an issue because the khronometric branch assumes the condition $\lambda\neq 0$.
Furthermore, notice that the shift and one of the vector Love numbers in Eqs.~\eqref{eq:tidalLove} and~\eqref{eq:tidalVectorLove} are independent of $\lambda$. Although $\tilde H_1$ in Eq.~\eqref{eq:H1_tilde} still depends on $\lambda$ and vanishes in the limit $\lambda \to 0$, such $\lambda$ dependence cancels out when computing the shift Love number. This is interesting because if these quantities were to be measured, we can conclude the existence of a non-GR effect. An exact method to measure these khronometric Love numbers remains unclear and is left to future work.

\subsubsection{Physical Interpretation}

We end this section by discussing the physical interpretation and potential observability of the new Love numbers. First, we note that these perturbations do not seem to be a gauge artefact. As we already mentioned, $\tilde H_1 = 0$ in GR. This is not the case in khronometric gravity as $\tilde H_1$ is coupled to the vector perturbation $V$, and these fields cannot be removed through a coordinate transformation. Namely, if we tried to eliminate $V$ from the vector field through a coordinate transformation, this degree of freedom would now appear in the metric, and vice versa.

If there are indeed nonvanishing tidal perturbations in $\tilde H_1$ and $V$, how would khronometric Love numbers enter observable quantities? One possibility is through gravitational waves from binary neutron star mergers.  The LIGO/Virgo Collaboration has measured the leading tidal Love number with GW170817~\cite{LIGOScientific:2017vwq,LIGOScientific:2018hze}. There are other subleading tidal Love numbers that are encoded in the waveform and may be extracted with future observations.  In GR, tidal perturbations in the $(t,A)$ component of the metric \kent{(with $A=(\theta,\phi)$)} are governed by magnetic-parity perturbations. Such perturbations give rise to magnetic tidal Love numbers. The leading quadrupolar magnetic Love number enters at 6th post-Newtonian order in the waveform (the leading electric-parity Love number enters at the 5th order)~\cite{Yagi:2013sva}. The khronometric Love numbers \sid{originating from $\tilde{H}_1$} may enter at the same order in the waveform, though a more detailed analysis needs to be done to determine whether they actually impact observables or not, and this is beyond the scope of this paper. \sid{Since $V$ originates from the vector field, it is unlikely that the Love numbers derived from $V$ will enter in the waveform as described above.}

\section{Conclusion and Discussion}\label{sec:conclusion}

We investigated the I-Love-Q universal relations for khronometric gravity in the limit $\abzero$, keeping only $\lambda$ free. We find that $\lambda$ is absent from the equations of motion for the background and at first order in slow rotation. When investigating second order solutions in slow rotation, we found that $\lambda$ is absent from the field variables that determine the quadrupole moment as well. Similarly, when we considered a weakly tidally-deformed perturbation, $\lambda$ does not enter the field equation components for the tidal Love number. Thus, we found that the moment of inertia, the tidal Love number, and the quadrupole moment of neutron stars in khronometric gravity match those of GR in the limit $\abzero$ for all equations of state. This means that the I-Love-Q relations cannot be used to test the coupling constant $\lambda$ in khronometric gravity, as they will be identical to those in GR. 

At second order in slow rotation and first order in tidal deformation, however, we did find seemingly nonvanishing $\lambda$ dependence that couples to two field functions, $V$ and $H_1$, which do not contribute to the calculation of I-Love-Q. We showed that when imposing regularity at the center and asymptotic flatness at spatial infinity, $V$ and $H_1$ must vanish for slowly-rotating neutron stars. We showed this by analyzing the field equations with nonvanishing $\lambda$ dependence in a PM framework. The field variables were expanded in powers of compactness, and the resulting solutions for $V$ and $H_1$ were found in both the exterior and interior of the star, where we imposed the appropriate boundary condition. We were able to inductively prove that the physical solution to this system is one where $V$ and $H_1$ vanish identically, indicating that there are no nonvanishing fields with $\lambda$ dependence. These results hold regardless of the equation of state and shows that slowly-rotating neutron stars in khronometric gravity are identical to those in GR in the $\abzero$ limit. On the other hand, the $\lambda$ dependence in $H_1$ and $V$ does not vanish for the tidally-deformed case, and one could introduce new shift and vector Love numbers as in Sec.~\ref{subsect:KLoveNumber}, though further studies are necessary to remove the ambiguity in its definition and to investigate its observability.

This work can be extended in a few other ways.
The reason why $\lambda$ only enters in the $(t,r)$ and $(t,\theta)$ components of the field equations at second order in rotation is because the metric and the vector field are functions of $(r,\theta)$ only. One possible future research direction is to consider time-dependent perturbations to extract oscillation frequencies of neutron stars. Then, $\lambda$ may enter in other components of the field equations and e.g. the fundamental-mode frequencies may depend on $\lambda$ (though black hole quasinormal modes were shown to be the same as in GR when $\abzero$~\cite{Franchini:2021bpt}). 

There are many new avenues to explore the additional Love numbers found above. For example, it would be interesting to compute similar Love numbers for black holes and use those as references to compute neutron star Love numbers as done in GR~\cite{Gralla:2017djj,Creci:2021rkz} to eliminate some of the ambiguities in their definition.
It would also be worthwhile to extend recent work that uses wave scattering to compute Love numbers as another way to define Love numbers free of ambiguities~\cite{Creci:2021rkz}.
Another potential avenue for future work is to study if a similar shift Love number appears in GR when electromagnetic fields are present. Tidal perturbations of Reisnner-Nordstr\"om black holes have been studied in~\cite{remie}. The authors found that $H_1 = 0$, and thus, there are no shift Love numbers for charged, non-rotating black holes. A similar analysis has been carried out for magnetized neutron stars in~\cite{Zhu:2020imp}, though the authors have focused on perturbations to the diagonal components of the metric. It would be interesting to extend their analysis to include perturbations to the off-diagonal components of the metric and the electromagnetic field.

One can easily apply the formulation presented in this paper to other Lorentz-violating theories of gravity, such as Einstein-\AE ther theory~\cite{Jacobson:2007veq}. The most up-to-date bounds (including those from binary pulsar and gravitational wave observations) on coupling constants in this theory have recently been derived in~\cite{Gupta:2021vdj}. The authors showed that there remains one constant associated to the vorticity of the vector field congruence~\cite{Jacobson:2013xta} that is unconstrained. It would be interesting to study how the I-Love-Q relations depend on this constant and whether one can use neutron star observations to probe Einstein-\AE ther theory through the I-Love-Q relations. 
Work along this direction is currently in progress~\cite{kai}.

\acknowledgments
We thank Eric Poisson, Philippe Landry, and Enrico Barausse for insightful discussions on vector Love numbers and khronometric gravity.
K.Y. acknowledges support from NSF Grant PHY-1806776, NASA Grant 80NSSC20K0523, a Sloan Foundation Research Fellowship and the Owens Family Foundation. 
K.Y. would like to also acknowledge support by the COST Action GWverse CA16104 and JSPS KAKENHI Grants No. JP17H06358. N.Y.~acknowledges support through NASA ATP Grant No. 17-ATP17-0225, No. NNX16AB98G and No. 80NSSC17M0041 and NSF Award No.~1759615.

\appendix
\sid{
\section{Khronometric Gravity as a Vector-Tensor or Scalar-Tensor Theory}
\label{app:scalarTensorAnalysis}
In this section, we take a closer look at how and why khronometric gravity can be regarded as either a vector-tensor or a scalar-tensor theory. We mentioned in the main text that the \ae ther equation of motion may be derived by varying the action given in Eq.~\eqref{eq:action} with respect to khronon scalar $T.$ This may lead one to conclude that the theory should be considered to be a scalar-tensor theory. However, one could equivalently derive the appropriate equations of motion by varying the Einstein-\ae ther action with respect to vector field $U_\mu$ while imposing five Lagrange multipliers~\cite{Yagi:2013ava}, four of the form $l_\alpha w^\alpha$ and one of the form $l_5 (U^\mu U_\mu-1)$, where $w^\mu$ is the vorticity vector defined in Eq.~\eqref{eq:vorticityDef}.}  \sid{Note the subtlety that one \emph{must} vary the action with respect to the covariant \ae ther vector $U_\mu$ before imposing the zero vorticity condition $w^\mu=0$ to get correct 
results~\cite{Barausse:2012qh}.} \sid{This equivalence is a consequence of the hypersurface orthogonality condition satisfied by the vector field in khronometric theory. This condition is the result of Frobenius' theorem, which is bijective~\cite{Poisson:2009pwt}. Hence, whether one starts with a scalar or a vector is on equivalent footing. However, the relationship between Einstein-\ae ther theory and khronometric gravity is more straightforward if one regards khronometric gravity as a vector-tensor theory, as we do in this work.}

\sid{To further illustrate this point, we shall show that one would reach the same conclusions as we do in the main body of this paper if one were to perturb the khronon scalar instead of the vector field. Additionally, from the zero-vorticity condition, we can explicitly prove why axial sector perturbations to the vector field must vanish, as we found above. To start, note that the khronon scalar $\tilde{T}$ is given by the initial time coordinate that we consider, taken to be the coordinate defined by the static Killing vector of the spacetime. When considering the perturbative effect of slow-rotation and tidal deformation, one must then perturb the khronon scalar. Let us denote this perturbed quantity as $T=\tilde{T}+\zeta\, \delta T(r,\theta,\phi)$, and let us form an ansatz for a separable solution using spherical harmonics, yielding
\be 
\delta T(r,\theta,\phi)=\tau(r)Y_{lm}(\theta,\phi).
\ee 
Here, $\zeta$ is a bookkeeping parameter for the order of the perturbation and $\tau$ is an arbitrary function of $r$.
Let us also establish the following notation:
\ba 
U_\mu&=&{}^0U_\mu+\zeta\,\delta U_\mu,\\
g_{\mu\nu}&=&\bgmetric_{\mu\nu}+\zeta\,\pertmetric_{\mu\nu},
\ea
where the superscript zero denotes the static background quantities and the delta terms denote perturbations. The goal here is to find $\delta U_\mu$ in terms of the other defined quantities.}

\sid{ Let us now derive the vector pertrubations from the khronon scalar perturbation. By linearizing Eq.~\eqref{eq:Ukhronon}, we find
\bw
\ba 
U_\mu&=&\frac{\partial_\mu T}{\sqrt{\partial_\nu T\partial^\nu T}}=\frac{\partial_\mu (\tilde{T}+\zeta\,\delta T)}{\sqrt{\partial_\nu (\tilde{T}+\zeta\,\delta T)\partial^\nu (\tilde{T}+\zeta\,\delta T)}}=\frac{\partial_\mu (\tilde{T}+\zeta\,\delta T)}{\sqrt{\partial_\nu \tilde{T}\partial^\nu \tilde{T}
+2\zeta\,\partial_\nu \tilde{T}\partial^\nu (\delta T)}}+\mathcal O(\zeta^2)\nonumber \\
&=&\frac{\partial_\mu (\tilde{T}+\zeta\,\delta T)}{\sqrt{\partial_\nu \tilde{T}\partial_\rho \tilde{T}(\bgmetric^{\nu\rho}+\zeta \, \delta g^{\nu\rho})
+2\zeta\,\partial_\nu \tilde{T}\partial_\rho (\delta T)\,\bgmetric^{\nu\rho}}}+\mathcal O(\zeta^2)\nonumber\\
&=&\frac{\partial_\mu (\tilde{T}+\zeta\,\delta T)}{\sqrt{\partial_\nu \tilde{T}\partial_\rho \tilde{T}\,\bgmetric^{\nu\rho}}}\left\{1-\frac \zeta 2\left[\partial_\nu \tilde{T}\partial_\rho \tilde{T}\, \delta g^{\nu\rho}
+2\partial_\nu \tilde{T}\partial_\rho (\delta T)\,\bgmetric^{\nu\rho}\right]\right\}+\mathcal O(\zeta^2)\nonumber\\
&=&\frac{\partial_\mu \tilde{T}}{\sqrt{\partial_\nu \tilde{T}\partial_\rho \tilde{T}\,\bgmetric^{\nu\rho}}}+\zeta\left\{\frac{\partial_\mu (\delta T)}{\sqrt{\partial_\nu \tilde{T}\partial_\rho \tilde{T}\,\bgmetric^{\nu\rho}}}- \frac{\partial_\mu \tilde{T}\left[\partial_\nu \tilde{T}\partial_\rho \tilde{T}\, \delta g^{\nu\rho}+2\partial_\nu \tilde{T}\partial_\rho (\delta T)\,\bgmetric^{\nu\rho}\right]}{2\sqrt{\partial_\nu \tilde{T}\partial_\rho \tilde{T}\,\bgmetric^{\nu\rho}}}\right\}+\mathcal O(\zeta^2)\nonumber\\
&=&{}^0U_\mu+\zeta\,\delta U_\mu+\mathcal O(\zeta^2),
\ea 
where we have successfully split the above into background and perturbed vector field quantities. We can simplify the above by noting that our spacetime and vector field are stationary, our background metric is diagonal, and our coordinates are adapted to $\tilde T=\tilde t.$ We then find
\ba 
\delta U_\mu&=&\frac{\delta^i_\mu\partial_i (\delta T)}{\sqrt{\partial_{\tilde t} \tilde{T}\partial_{\tilde t} \tilde{T}\,\bgmetric^{{\tilde t}{\tilde t}}}}- \frac{\delta^{\tilde t}_\mu\partial_{\tilde{t}} \tilde{T}\left[\partial_{\tilde{t}} \tilde{T}\partial_{\tilde{t}} \tilde{T}\, \delta g^{\tilde{t}\tilde{t}}+2\partial_{\tilde{t}} \tilde{T}\partial_i (\delta T)\,\bgmetric^{\tilde{t} i}\right]}{2\sqrt{\partial_{\tilde{t}} \tilde{T}\partial_{\tilde{t}} \tilde{T}\,\bgmetric^{\tilde{t}\tilde{t}}}}=\frac{1}{{\sqrt{\bgmetric^{{\tilde t}{\tilde t}}}}}\left[\delta^i_\mu\partial_i (\delta T)- \frac{\delta^{\tilde t}_\mu \delta g^{\tilde{t}\tilde{t}}}{2}\right].
\ea 
\ew
We can thus see that by plugging in the metric and khronon perturbation ansatz above while taking $\zeta=\varepsilon^2$ and $\tau(r)=\kappa\, V(r)$, we recover Eq.~\eqref{eq:generalAEtherVorticityFree}}\sid{. We note that the coordinate transformation taken in Eq.~\eqref{eq:coordinateTransformation} is essentially taking the perturbed khronon scalar to be our new time coordinate.}

\sid{
The above indicates that odd-parity perturbations are eliminated by the vorticity free condition, which we will explicitly prove here. Consider an odd-parity perturbation, which has the form 
\be 
\delta U_\mu=\tau(r)\delta_{\mu }{}^A\varepsilon_A{}^{B}\partial_B [Y_{lm}(\theta,\phi)],
\ee
where capital Latin letters indicate coordinates on the 2-sphere and $\varepsilon_{AB}$ is the corresponding 2-D Levi-Civita tensor. Now, keeping in mind that the vector field is stationary and that its background value is static and solely a function of $r$, we find that the vorticity-free condition at first order in perturbation gives the following constraints:
\bw 
\ba 
\delta w^\mu=0&=&\varepsilon^{\mu\nu\rho\sigma}\left[\delta U_\nu \partial_\rho U_\sigma+U_\nu \partial_\rho(\delta U_\sigma)\right]+\delta\varepsilon^{\mu\nu\rho\sigma} U_\nu \partial_\rho U_\sigma=\varepsilon^{\mu\nu\rho\sigma}\left[\delta U_\nu \partial_\rho U_\sigma-U_\sigma \partial_\rho(\delta U_\nu)\right]+\delta\varepsilon^{\mu \tilde{t} i \tilde{t}} U_{\tilde{t} } \partial_i U_{\tilde{t} }\nonumber\\
&=&\varepsilon^{\mu A i \tilde{t}}\left[\delta U_A \partial_i U_{\tilde{t}}-U_{\tilde{t}} \partial_i(\delta U_A)\right]=\delta^\mu_B\varepsilon^{B A r \tilde{t}}\delta U_A \partial_r U_{\tilde{t}}-\varepsilon^{\mu A i \tilde{t}}U_{\tilde{t}} \partial_i(\delta U_A)\nonumber\\
&=&\delta^\mu_B\varepsilon^{B A r \tilde{t}}(\delta U_A \partial_rU_{\tilde{t}}-U_{\tilde{t}}\partial_r \delta U_A)-\delta^\mu_r\varepsilon^{rAB\tilde{t}}U_{\tilde{t}}\partial_B \delta U_A.
\ea
Now consider the $\mu=r$ component of this equation. Let  the metric on the 2-sphere be $h_{AB}$ and the Levi-Civita symbol be $\bar\varepsilon_{AB}=\frac{1}{\sqrt{h}}\varepsilon_{AB}=\frac{1}{\sin\theta}\,\varepsilon_{AB}$ and $\bar\varepsilon^{AB}=\sqrt{h}\,\varepsilon^{AB}=\sin\theta\,\varepsilon^{AB}$, where $h$ is the 2-sphere metric determinant. We then find that 
\ba 
\delta w^r&\propto&\varepsilon^{AB}\partial_B \delta U_A=\tau(r)\varepsilon^{AB}\partial_B[h^{DC}\varepsilon_{AD} \partial_C Y_{lm}(\theta,\phi)]=\frac{\tau(r)}{\sqrt{h}}\varepsilon^{AB}\varepsilon_{AD}\partial_B[\sqrt{h}\partial^D Y_{lm}(\theta,\phi)]=\nonumber\\
&=&\frac{\tau(r)}{\sin\theta}\delta^{B}_D\partial_B[ \sin\theta\,\partial^D Y_{lm}(\theta,\phi)]=\frac{\tau(r)}{\sin\theta}\partial_D[ \sin\theta\,\partial^D Y_{lm}(\theta,\phi)]=-\tau(r)l(l+1)Y_{lm}(\theta,\phi)=0,
\ea 
where we have used the Laplace equation and the fact that $l$ is nonzero in the last line.
\ew
From this, we can conclude that $\tau(r)=0$, and thus $\delta U_\mu=0$ for odd-parity vector perturbations. We see this more general analysis supports the results we found from Eq.~\eqref{eq:generalVorticity}.
}
\section{Field Equations for Neutron Stars}\label{sec:AppendixUncontrainedValues}
Here, we present the field equations for neutron stars in khronometric gravity. We first present equations for background neutron stars with no rotation and tidal deformation. We then present equations for slowly-rotating neutron stars at first and second orders in spin, followed by weakly tidally-deformed neutron stars at first order in tidal perturbation.

\subsection{Background}\label{subsect:backgroundEqns}

We first calculate and present the field equations at $\mathcal{O}(\varepsilon^0)$, which  serve as background equations.
After inserting the metric and vector ansatz, given by taking $\varepsilon \to 0$ in Eqs.~\eqref{eq:generalMetric} and~\eqref{eq:generalAEther}, into Eq.~\eqref{eq:EinsteinEqnsTensor}, we can use $E_{tt}=0$ and $E_{rr}=0$ along with the matter equation of motion $\nabla^\mu T^\mathrm{(mat)}_{\mu r}=0$ to find
\bw
\ba\label{eq:dMdr}
\frac{dM}{dr}&=&\frac{\left(4-\alpha\right) M+2 \left[\sqrt{(r-2 M)
   \left(r -2 M + \alpha  M+4 \pi  \alpha  p_0 r^3\right)}-r\right]}{\alpha r}+\frac{8 \pi r^2  [\left(2\alpha
   -1\right)   p_0 +   \epsilon_0 ]}{(2-\alpha)},\nonumber\\
\\\label{eq:dnudr}
\frac{d\nu}{dr}&=&4\frac{ \sqrt{(r-2 M) \left(r -2 M + \alpha  M+4 \pi  \alpha  p_0 r^3\right)}+2 M- r}{\alpha
    r (r-2 M)},\\ \label{eq:dpdr}
\frac{dp}{dr}&=&-2\frac{ (p_0+\epsilon_0 ) \left[\sqrt{(r-2 M) \left(r -2 M + \alpha  M+4 \pi  \alpha  p_0 r^3\right)}+2 M-r\right]}{\alpha  r (r-2M)}.
\ea
\ew
One may think that
the above equations diverge in the GR limit ($\alpha \to 0$), but if we expand them about $\alpha=0$, the equations become Eqs.~(142)--(144) of~\cite{Yagi:2013ava} (with $c_{14}$ replaced by $\alpha$) and they correctly reduce to the GR TOV equations in the $\alpha \to 0$ limit.

\subsection{First Order in Rotation}

Next, we derive an equation for slowly-rotating neutron stars at first order in spin.
Keeping up to $\mathcal{O}(\varepsilon)$ terms in Eqs.~\eqref{eq:generalMetric} and~\eqref{eq:generalAEther}, 
we find that the only nontrivial first-order field equation is obtained from $E_{t\phi}=0$, which reduces to 
\bw
\ba
\label{eq:spin1}
\frac{d^2\omega}{dr^2}&=&\left\{\frac{8 \pi  r^2
   [(2 \alpha -1) p_0 +\epsilon_0 ]}{(2-\alpha)  (r-2 M)}+4\frac{ \sqrt{(r-2 M)
   \left[(\alpha -2) M+4 \pi  \alpha  p_0 r^3+r\right]}- (\alpha+1)r+2(3\alpha/4+1)M}{\alpha   r (r-2 M)}\right\}\frac{d\omega}{dr}\nonumber\\&&+\frac{16 \pi  r (p_0+\epsilon_0 )}{(1-\beta) (r-2 M)}\omega.
\ea
\ew
Since this expression is independent of $\lambda,$ the equation coincides with that in GR to linear order in $\varepsilon$ after taking the $\abzero$ limit. 

\subsection{Second Order in Rotation}
\label{app:2nd}
We now move on to deriving modified field equations at second order in rotation. Let us first derive the coupled equations for $H_0$, $K$ and $H_2$. From the $r$ and $\theta$ components of the matter equation of motion, we solve for $\epsilon_2$ and $p_2$, respectively. However, we note that $\epsilon_2$ does not enter in the calculation of the quadrupole moment since it appears in $E_{tt}=0,$ which is not used to derive the following coupled differential equations. We then use $E_{\theta\theta} -E_{\phi\phi} =0$ to find 
\bw
\be\label{eq:H2_rot}
H_2=H_0+\frac{r^3e^{-\nu}}{3}
\left[(\beta-1)(r-2M)\left(\frac{d\omega}{dr}\right)^2-16\pi r(p_0+\epsilon_0)\omega^2\right],
\ee
Next, using $E_{r\theta}=0$ and $E_{rr}=0$, we find
\ba\label{eq:dKdr_rot}
\frac{dK}{dr}&=&\frac{dH_0}{dr}+\left[\frac{(\alpha-2)}{\alpha r}-2\frac{(\alpha-1)\sqrt{4\alpha\pi p_0 r^3+\alpha M+r-2M}}{\alpha r \sqrt{r-2M}}\right]H_0+\left[\frac{(\alpha-2)}{\alpha r}+\frac{2\sqrt{4\alpha\pi p_0 r^3+\alpha M+r-2M}}{\alpha r \sqrt{r-2M}}\right]H_2,\nonumber\\ 
&& \\ \label{eq:dH0dr_rot}
\frac{dH_0}{dr}&=&\frac{1}{\alpha\sqrt{(r-2M)(4\alpha\pi p_0 r^3+\alpha M+r-2M)}} \Bigg\{\left[(\alpha-2)(r-2M)+\sqrt{(r-2M)(4\alpha\pi p_0 r^3+\alpha M+r-2M)}\right]\frac{dK}{dr}\nonumber\\ &-&2\alpha K-\alpha\left[4\pi r^2\right(p_0+\epsilon_0)-3]H_0+(8\pi r^2p_0+1)H_2-\frac{\alpha e^{-\nu}r^3}{6}\left[(\beta-1)(r-2M)\left(\frac{d\omega}{dr}\right)^2+16\pi r \omega^2(p_0+\epsilon_0)\right]\Bigg\}.\nonumber\\
\ea
\ew

Next, we derive the coupled equations for $H_1$ and $V$. 
From $E_{tr}=0$ and $E_{t\theta}=0$,
we find 
\begin{equation}\label{eq:H1_generic}
   \frac{d^2H_1}{dr^2} = \frac{1}{b_0}\left(b_1 \frac{dH_1}{dr} + b_2 H_1+b_3 \frac{dV}{dr} + b_4 V\right),
\end{equation} 
with
\bw
\ba
b_0&=&\alpha^2(\alpha-2)(\beta-\lambda)r^2(r-2M)\sqrt{(4\pi\alpha r^3p_0+\alpha M-2M+r)(r-2M)},\\
b_1&=&-\alpha (\beta+\lambda)r\{2(\alpha-2)(\alpha-8)M+24\pi\alpha r^3[(2\alpha-1)p_0+\epsilon_0]+2(\alpha+4)(\alpha-2)r\}\nonumber\\&\times&\sqrt{(4\pi\alpha r^3p_0+\alpha M-2M+r)(r-2M)}
+8(4\pi\alpha r^3p_0+\alpha M-2M+r)[\alpha(\alpha-2)(\beta+\lambda)r(r-2M)],\\
b_2&=&-\left\{8\pi\alpha^2(\beta+\lambda)r^4\frac{d\epsilon_0}{dr}+2[(\beta+3\lambda+2)\alpha^2+2(7\beta-4+3\lambda)\alpha-24(\beta+\lambda)](\alpha-2)M\right.\nonumber\\ &+&8\pi \alpha r^3[(2\beta+6\lambda+5)\alpha^2+2(10\beta+9\lambda-2)\alpha-18(\beta+\lambda)]p_0+8\pi\alpha r^3 [\alpha^2+2\alpha(3\beta+4\lambda)+2(\beta+\lambda)]\epsilon_0\nonumber\\  &-& \left.(\alpha-2)[(5\beta+2\lambda-3)\alpha^2+4(\beta-\lambda-2)\alpha-24(\beta+\lambda)]r\right\}\sqrt{(4\pi\alpha r^3p_0+\alpha M-2M+r)(r-2M)}\nonumber\\&+&(4\pi\alpha r^3p_0+\alpha M-2M+r)\left\{16[(\beta-2)\alpha -3(\beta+\lambda)](\alpha-2)M+48\pi\alpha r^3(\beta+\lambda)(2\alpha-1)p_0\right.\nonumber\\&+& \left.16\pi\alpha r^3(\beta+\lambda)(2\alpha-1)\epsilon_0-[(\beta-\lambda-2)\alpha-6(\beta+\lambda)](\alpha-2)r\right\}, \\
b_3&=&-3\alpha^2(\alpha-2)(\beta+2\lambda+1)r\,e^\nu\sqrt{(4\pi\alpha r^3p_0+\alpha M-2M+r)(r-2M)},\\
b_4&=&e^\nu(\alpha-2)(\beta+\lambda)[12(\alpha+1)\sqrt{(4\pi\alpha r^3p_0+\alpha M-2M+r)(r-2M)}-12\alpha(4\pi\alpha r^3p_0+\alpha M-2M+r)],
\ea
and 
\begin{equation}\label{eq:V_generic}
   \frac{d^2V}{dr^2} =- \frac{1}{c_0}\left(c_1 \frac{dV}{dr} + c_2 V+c_3 \frac{dH_1}{dr} + c_4 H_1\right),
\end{equation}
with 
\ba
c_0&=&\alpha(\alpha-2)(\beta-1)r(r-2M)\sqrt{(4\pi\alpha r^3p_0+\alpha M-2M+r)(r-2M)},\\
c_1&=&(\beta-1)(\alpha-2)(r-2M)(4\pi\alpha r^3p_0+\alpha M-2M+r)+2(\beta-1)\sqrt{(4\pi\alpha r^3p_0+\alpha M-2M+r)(r-2M)}\nonumber\\ &\times&\{(\alpha^2+2\alpha-8)M+4\pi\alpha r^3[(2\alpha-1)p_0+\epsilon_0]-2\alpha r+4r\},\\
c_2&=&-4\alpha\{4\pi r^2[(\{2\beta-3\}\alpha+\beta)p_0+(2\beta-\alpha)\epsilon_0]+3(\alpha-2)(\beta+\lambda)\}\sqrt{(4\pi\alpha r^3p_0+\alpha M-2M+r)(r-2M)},\\
c_3&=&-\alpha(\alpha-2)(\beta+2\lambda+1)r(r-2M)\sqrt{(4\pi\alpha r^3p_0+\alpha M-2M+r)(r-2M)},\\ \label{eq:endOfVEq_generic}
c_4&=&e^{-\nu}(\beta+\lambda)(\alpha-2)(r-2M)(4\pi\alpha r^3p_0+\alpha M-2M+r)+e^{-\nu}\sqrt{(4\pi\alpha r^3p_0+\alpha M-2M+r)(r-2M)}\nonumber\\&\times&\{2(\alpha-2)[(3\beta+2\lambda-1)\alpha+4\beta+4\lambda]M-8\pi\alpha r^3(\beta+2\lambda+1)[(2\alpha-1)p_0+
\epsilon_0]-4r(\alpha+1)(\alpha-2)(\beta+\lambda)\}.\nonumber\\
\ea
\ew
These equations reduce to Eqs.~\eqref{eq:d2H1dr2} and~\eqref{eq:d2Vdr2} after taking the limit $\abzero$.

\subsection{First Order in Tidal Deformation}
\label{sec:tidal_main}

Let us now present the field equations for even-parity tidal perturbation at first order. As explained in Sec.~\ref{sec:tidal}, we can use the results at second order in spin by setting the contribution at $\mathcal{O}(\epsilon)$ to 0. Setting $\omega = 0$ and $d\omega/dr=0$ in Eqs.~\eqref{eq:H2_rot}--\eqref{eq:dH0dr_rot}, we find $H_0 = H_2 \equiv H$ and a system of coupled equations for $H$ and $K$. One can further eliminate $K$ from this equation to find
\be\label{eq:H_generic}
   \frac{d^2H}{dr^2} = \frac{1}{a_0}\left(a_1 \frac{dH}{dr} + a_2 H\right),
\ee
with
\bw
\ba
a_0&=& \alpha^2(r-2M)(\alpha-2)r^2\left\{r+\alpha \pi r^3 p_0+(\alpha-2)M-\sqrt { \left( r-2M\right)  \left[ 4
\pi\alpha \, r^3\ p_0+M (\alpha -2)+r \right] }\right\},\\
a_1&=& 2\alpha^2r\sqrt { \left( r-2M\right)  \left[ 4
\pi\alpha \, r^3\ p_0+M (\alpha -2)+r \right] }\left(r\left[4\pi r^2\left\{\epsilon_0+(2\alpha-1)p_0\right\}+\alpha-2\right]-M(\alpha-2)\right)\nonumber \\ &-&2\alpha^2r\left[r+\left(\alpha-2\right)M+4\pi\alpha r^3\, p_0\right]\left(r\left\{4(2\alpha-1)\pi r^2p_0+\alpha-2+4\pi r^2 \epsilon_0 \right\} - (\alpha-2) M \right),
\\
a_2&=& 4\sqrt { \left( r-2M\right)  \left[ 4
\pi\alpha \, r^3\ p_0+M (\alpha -2)+r \right] }\nonumber\\ &\times& \left(2(\alpha-2)^2\left(3\alpha-8\right)M+r\left\{6\alpha\left(4\alpha^2-19\alpha+16\right)\pi r^2 p_0-10\alpha^2\pi r^2 \epsilon_0 -\frac{3\alpha^3}{2}+11\alpha^2-32\alpha+32\right\}-\alpha^3\pi r^4 \frac{d\epsilon_0}{dr}\right)\nonumber\\ &-&2\left(r\left\{64-64\alpha+4\alpha\pi r^2\left[(4\alpha^2-25\alpha+16)p_0-5\alpha\epsilon_0\right]-3\alpha^3+22\alpha^2\right\}+4(\alpha-8)(\alpha-2)^2M\right)
\nonumber\\ \label{eq:H_generic_end}&\times&\left[r+\left(\alpha-2\right)M+4\pi\alpha r^3\, p_0\right].
\ea
\ew
Notice that the above equation only depends on $\alpha$, and reduces to the GR case when taking $\alpha\to0 $ as stated before. 

\bibliography{bibliography.bib}

\begin{thebibliography}{74}%
\makeatletter
\providecommand \@ifxundefined [1]{%
 \@ifx{#1\undefined}
}%
\providecommand \@ifnum [1]{%
 \ifnum #1\expandafter \@firstoftwo
 \else \expandafter \@secondoftwo
 \fi
}%
\providecommand \@ifx [1]{%
 \ifx #1\expandafter \@firstoftwo
 \else \expandafter \@secondoftwo
 \fi
}%
\providecommand \natexlab [1]{#1}%
\providecommand \enquote  [1]{``#1''}%
\providecommand \bibnamefont  [1]{#1}%
\providecommand \bibfnamefont [1]{#1}%
\providecommand \citenamefont [1]{#1}%
\providecommand \href@noop [0]{\@secondoftwo}%
\providecommand \href [0]{\begingroup \@sanitize@url \@href}%
\providecommand \@href[1]{\@@startlink{#1}\@@href}%
\providecommand \@@href[1]{\endgroup#1\@@endlink}%
\providecommand \@sanitize@url [0]{\catcode `\\12\catcode `\$12\catcode
  `\&12\catcode `\#12\catcode `\^12\catcode `\_12\catcode `\%12\relax}%
\providecommand \@@startlink[1]{}%
\providecommand \@@endlink[0]{}%
\providecommand \url  [0]{\begingroup\@sanitize@url \@url }%
\providecommand \@url [1]{\endgroup\@href {#1}{\urlprefix }}%
\providecommand \urlprefix  [0]{URL }%
\providecommand \Eprint [0]{\href }%
\providecommand \doibase [0]{https://doi.org/}%
\providecommand \selectlanguage [0]{\@gobble}%
\providecommand \bibinfo  [0]{\@secondoftwo}%
\providecommand \bibfield  [0]{\@secondoftwo}%
\providecommand \translation [1]{[#1]}%
\providecommand \BibitemOpen [0]{}%
\providecommand \bibitemStop [0]{}%
\providecommand \bibitemNoStop [0]{.\EOS\space}%
\providecommand \EOS [0]{\spacefactor3000\relax}%
\providecommand \BibitemShut  [1]{\csname bibitem#1\endcsname}%
\let\auto@bib@innerbib\@empty
\bibitem [{\citenamefont {Will}(2018)}]{Will:2018bme}%
  \BibitemOpen
  \bibfield  {author} {\bibinfo {author} {\bibfnamefont {C.~M.}\ \bibnamefont
  {Will}},\ }\href@noop {} {\emph {\bibinfo {title} {{Theory and Experiment in
  Gravitational Physics}}}}\ (\bibinfo  {publisher} {Cambridge University
  Press},\ \bibinfo {year} {2018})\BibitemShut {NoStop}%
\bibitem [{\citenamefont {Will}(2014)}]{Will:2014kxa}%
  \BibitemOpen
  \bibfield  {author} {\bibinfo {author} {\bibfnamefont {C.~M.}\ \bibnamefont
  {Will}},\ }\bibfield  {title} {\bibinfo {title} {{The Confrontation between
  General Relativity and Experiment}},\ }\href
  {https://doi.org/10.12942/lrr-2014-4} {\bibfield  {journal} {\bibinfo
  {journal} {Living Rev. Rel.}\ }\textbf {\bibinfo {volume} {17}},\ \bibinfo
  {pages} {4} (\bibinfo {year} {2014})},\ \Eprint
  {https://arxiv.org/abs/1403.7377} {arXiv:1403.7377 [gr-qc]} \BibitemShut
  {NoStop}%
\bibitem [{\citenamefont {Berti}\ \emph {et~al.}(2015)\citenamefont {Berti}
  \emph {et~al.}}]{Berti:2015itd}%
  \BibitemOpen
  \bibfield  {author} {\bibinfo {author} {\bibfnamefont {E.}~\bibnamefont
  {Berti}} \emph {et~al.},\ }\bibfield  {title} {\bibinfo {title} {{Testing
  General Relativity with Present and Future Astrophysical Observations}},\
  }\href {https://doi.org/10.1088/0264-9381/32/24/243001} {\bibfield  {journal}
  {\bibinfo  {journal} {Class. Quant. Grav.}\ }\textbf {\bibinfo {volume}
  {32}},\ \bibinfo {pages} {243001} (\bibinfo {year} {2015})},\ \Eprint
  {https://arxiv.org/abs/1501.07274} {arXiv:1501.07274 [gr-qc]} \BibitemShut
  {NoStop}%
\bibitem [{\citenamefont {Jain}\ and\ \citenamefont
  {Khoury}(2010)}]{Jain:2010ka}%
  \BibitemOpen
  \bibfield  {author} {\bibinfo {author} {\bibfnamefont {B.}~\bibnamefont
  {Jain}}\ and\ \bibinfo {author} {\bibfnamefont {J.}~\bibnamefont {Khoury}},\
  }\bibfield  {title} {\bibinfo {title} {{Cosmological Tests of Gravity}},\
  }\href {https://doi.org/10.1016/j.aop.2010.04.002} {\bibfield  {journal}
  {\bibinfo  {journal} {Annals Phys.}\ }\textbf {\bibinfo {volume} {325}},\
  \bibinfo {pages} {1479} (\bibinfo {year} {2010})},\ \Eprint
  {https://arxiv.org/abs/1004.3294} {arXiv:1004.3294 [astro-ph.CO]}
  \BibitemShut {NoStop}%
\bibitem [{\citenamefont {Clifton}\ \emph {et~al.}(2012)\citenamefont
  {Clifton}, \citenamefont {Ferreira}, \citenamefont {Padilla},\ and\
  \citenamefont {Skordis}}]{Clifton:2011jh}%
  \BibitemOpen
  \bibfield  {author} {\bibinfo {author} {\bibfnamefont {T.}~\bibnamefont
  {Clifton}}, \bibinfo {author} {\bibfnamefont {P.~G.}\ \bibnamefont
  {Ferreira}}, \bibinfo {author} {\bibfnamefont {A.}~\bibnamefont {Padilla}},\
  and\ \bibinfo {author} {\bibfnamefont {C.}~\bibnamefont {Skordis}},\
  }\bibfield  {title} {\bibinfo {title} {{Modified Gravity and Cosmology}},\
  }\href {https://doi.org/10.1016/j.physrep.2012.01.001} {\bibfield  {journal}
  {\bibinfo  {journal} {Phys. Rept.}\ }\textbf {\bibinfo {volume} {513}},\
  \bibinfo {pages} {1} (\bibinfo {year} {2012})},\ \Eprint
  {https://arxiv.org/abs/1106.2476} {arXiv:1106.2476 [astro-ph.CO]}
  \BibitemShut {NoStop}%
\bibitem [{\citenamefont {Joyce}\ \emph {et~al.}(2015)\citenamefont {Joyce},
  \citenamefont {Jain}, \citenamefont {Khoury},\ and\ \citenamefont
  {Trodden}}]{Joyce:2014kja}%
  \BibitemOpen
  \bibfield  {author} {\bibinfo {author} {\bibfnamefont {A.}~\bibnamefont
  {Joyce}}, \bibinfo {author} {\bibfnamefont {B.}~\bibnamefont {Jain}},
  \bibinfo {author} {\bibfnamefont {J.}~\bibnamefont {Khoury}},\ and\ \bibinfo
  {author} {\bibfnamefont {M.}~\bibnamefont {Trodden}},\ }\bibfield  {title}
  {\bibinfo {title} {{Beyond the Cosmological Standard Model}},\ }\href
  {https://doi.org/10.1016/j.physrep.2014.12.002} {\bibfield  {journal}
  {\bibinfo  {journal} {Phys. Rept.}\ }\textbf {\bibinfo {volume} {568}},\
  \bibinfo {pages} {1} (\bibinfo {year} {2015})},\ \Eprint
  {https://arxiv.org/abs/1407.0059} {arXiv:1407.0059 [astro-ph.CO]}
  \BibitemShut {NoStop}%
\bibitem [{\citenamefont {Koyama}(2016)}]{Koyama:2015vza}%
  \BibitemOpen
  \bibfield  {author} {\bibinfo {author} {\bibfnamefont {K.}~\bibnamefont
  {Koyama}},\ }\bibfield  {title} {\bibinfo {title} {{Cosmological Tests of
  Modified Gravity}},\ }\href {https://doi.org/10.1088/0034-4885/79/4/046902}
  {\bibfield  {journal} {\bibinfo  {journal} {Rept. Prog. Phys.}\ }\textbf
  {\bibinfo {volume} {79}},\ \bibinfo {pages} {046902} (\bibinfo {year}
  {2016})},\ \Eprint {https://arxiv.org/abs/1504.04623} {arXiv:1504.04623
  [astro-ph.CO]} \BibitemShut {NoStop}%
\bibitem [{\citenamefont {Horava}(2009)}]{Horava:2009uw}%
  \BibitemOpen
  \bibfield  {author} {\bibinfo {author} {\bibfnamefont {P.}~\bibnamefont
  {Horava}},\ }\bibfield  {title} {\bibinfo {title} {{Quantum Gravity at a
  Lifshitz Point}},\ }\href {https://doi.org/10.1103/PhysRevD.79.084008}
  {\bibfield  {journal} {\bibinfo  {journal} {Phys. Rev. D}\ }\textbf {\bibinfo
  {volume} {79}},\ \bibinfo {pages} {084008} (\bibinfo {year} {2009})},\
  \Eprint {https://arxiv.org/abs/0901.3775} {arXiv:0901.3775 [hep-th]}
  \BibitemShut {NoStop}%
\bibitem [{\citenamefont {Blas}\ \emph {et~al.}(2011)\citenamefont {Blas},
  \citenamefont {Pujolas},\ and\ \citenamefont {Sibiryakov}}]{Blas:2010hb}%
  \BibitemOpen
  \bibfield  {author} {\bibinfo {author} {\bibfnamefont {D.}~\bibnamefont
  {Blas}}, \bibinfo {author} {\bibfnamefont {O.}~\bibnamefont {Pujolas}},\ and\
  \bibinfo {author} {\bibfnamefont {S.}~\bibnamefont {Sibiryakov}},\ }\bibfield
   {title} {\bibinfo {title} {{Models of non-relativistic quantum gravity: The
  Good, the bad and the healthy}},\ }\href
  {https://doi.org/10.1007/JHEP04(2011)018} {\bibfield  {journal} {\bibinfo
  {journal} {JHEP}\ }\textbf {\bibinfo {volume} {04}},\ \bibinfo {pages}
  {018}},\ \Eprint {https://arxiv.org/abs/1007.3503} {arXiv:1007.3503 [hep-th]}
  \BibitemShut {NoStop}%
\bibitem [{\citenamefont {Kostelecky}(2004)}]{Kostelecky:2003fs}%
  \BibitemOpen
  \bibfield  {author} {\bibinfo {author} {\bibfnamefont {V.~A.}\ \bibnamefont
  {Kostelecky}},\ }\bibfield  {title} {\bibinfo {title} {{Gravity, Lorentz
  violation, and the standard model}},\ }\href
  {https://doi.org/10.1103/PhysRevD.69.105009} {\bibfield  {journal} {\bibinfo
  {journal} {Phys. Rev. D}\ }\textbf {\bibinfo {volume} {69}},\ \bibinfo
  {pages} {105009} (\bibinfo {year} {2004})},\ \Eprint
  {https://arxiv.org/abs/hep-th/0312310} {arXiv:hep-th/0312310} \BibitemShut
  {NoStop}%
\bibitem [{\citenamefont {Mattingly}(2005)}]{Mattingly:2005re}%
  \BibitemOpen
  \bibfield  {author} {\bibinfo {author} {\bibfnamefont {D.}~\bibnamefont
  {Mattingly}},\ }\bibfield  {title} {\bibinfo {title} {{Modern tests of
  Lorentz invariance}},\ }\href {https://doi.org/10.12942/lrr-2005-5}
  {\bibfield  {journal} {\bibinfo  {journal} {Living Rev. Rel.}\ }\textbf
  {\bibinfo {volume} {8}},\ \bibinfo {pages} {5} (\bibinfo {year} {2005})},\
  \Eprint {https://arxiv.org/abs/gr-qc/0502097} {arXiv:gr-qc/0502097}
  \BibitemShut {NoStop}%
\bibitem [{\citenamefont {Jacobson}\ \emph {et~al.}(2006)\citenamefont
  {Jacobson}, \citenamefont {Liberati},\ and\ \citenamefont
  {Mattingly}}]{Jacobson:2005bg}%
  \BibitemOpen
  \bibfield  {author} {\bibinfo {author} {\bibfnamefont {T.}~\bibnamefont
  {Jacobson}}, \bibinfo {author} {\bibfnamefont {S.}~\bibnamefont {Liberati}},\
  and\ \bibinfo {author} {\bibfnamefont {D.}~\bibnamefont {Mattingly}},\
  }\bibfield  {title} {\bibinfo {title} {{Lorentz violation at high energy:
  Concepts, phenomena and astrophysical constraints}},\ }\href
  {https://doi.org/10.1016/j.aop.2005.06.004} {\bibfield  {journal} {\bibinfo
  {journal} {Annals Phys.}\ }\textbf {\bibinfo {volume} {321}},\ \bibinfo
  {pages} {150} (\bibinfo {year} {2006})},\ \Eprint
  {https://arxiv.org/abs/astro-ph/0505267} {arXiv:astro-ph/0505267}
  \BibitemShut {NoStop}%
\bibitem [{\citenamefont {Liberati}(2013)}]{Liberati:2013xla}%
  \BibitemOpen
  \bibfield  {author} {\bibinfo {author} {\bibfnamefont {S.}~\bibnamefont
  {Liberati}},\ }\bibfield  {title} {\bibinfo {title} {{Tests of Lorentz
  invariance: a 2013 update}},\ }\href
  {https://doi.org/10.1088/0264-9381/30/13/133001} {\bibfield  {journal}
  {\bibinfo  {journal} {Class. Quant. Grav.}\ }\textbf {\bibinfo {volume}
  {30}},\ \bibinfo {pages} {133001} (\bibinfo {year} {2013})},\ \Eprint
  {https://arxiv.org/abs/1304.5795} {arXiv:1304.5795 [gr-qc]} \BibitemShut
  {NoStop}%
\bibitem [{\citenamefont {Yagi}\ \emph
  {et~al.}(2014{\natexlab{a}})\citenamefont {Yagi}, \citenamefont {Blas},
  \citenamefont {Barausse},\ and\ \citenamefont {Yunes}}]{Yagi:2013ava}%
  \BibitemOpen
  \bibfield  {author} {\bibinfo {author} {\bibfnamefont {K.}~\bibnamefont
  {Yagi}}, \bibinfo {author} {\bibfnamefont {D.}~\bibnamefont {Blas}}, \bibinfo
  {author} {\bibfnamefont {E.}~\bibnamefont {Barausse}},\ and\ \bibinfo
  {author} {\bibfnamefont {N.}~\bibnamefont {Yunes}},\ }\bibfield  {title}
  {\bibinfo {title} {{Constraints on Einstein-\AE{}ther theory and Ho\v{r}ava
  gravity from binary pulsar observations}},\ }\href
  {https://doi.org/10.1103/PhysRevD.89.084067} {\bibfield  {journal} {\bibinfo
  {journal} {Phys. Rev. D}\ }\textbf {\bibinfo {volume} {89}},\ \bibinfo
  {pages} {084067} (\bibinfo {year} {2014}{\natexlab{a}})},\ \bibinfo {note}
  {[Erratum: Phys.Rev.D 90, 069902 (2014), Erratum: Phys.Rev.D 90, 069901
  (2014)]},\ \Eprint {https://arxiv.org/abs/1311.7144} {arXiv:1311.7144
  [gr-qc]} \BibitemShut {NoStop}%
\bibitem [{\citenamefont {Yagi}\ \emph
  {et~al.}(2014{\natexlab{b}})\citenamefont {Yagi}, \citenamefont {Blas},
  \citenamefont {Yunes},\ and\ \citenamefont {Barausse}}]{Yagi:2013qpa}%
  \BibitemOpen
  \bibfield  {author} {\bibinfo {author} {\bibfnamefont {K.}~\bibnamefont
  {Yagi}}, \bibinfo {author} {\bibfnamefont {D.}~\bibnamefont {Blas}}, \bibinfo
  {author} {\bibfnamefont {N.}~\bibnamefont {Yunes}},\ and\ \bibinfo {author}
  {\bibfnamefont {E.}~\bibnamefont {Barausse}},\ }\bibfield  {title} {\bibinfo
  {title} {{Strong Binary Pulsar Constraints on Lorentz Violation in
  Gravity}},\ }\href {https://doi.org/10.1103/PhysRevLett.112.161101}
  {\bibfield  {journal} {\bibinfo  {journal} {Phys. Rev. Lett.}\ }\textbf
  {\bibinfo {volume} {112}},\ \bibinfo {pages} {161101} (\bibinfo {year}
  {2014}{\natexlab{b}})},\ \Eprint {https://arxiv.org/abs/1307.6219}
  {arXiv:1307.6219 [gr-qc]} \BibitemShut {NoStop}%
\bibitem [{\citenamefont {Emir~Gümrükçüoğlu}\ \emph
  {et~al.}(2018)\citenamefont {Emir~Gümrükçüoğlu}, \citenamefont
  {Saravani},\ and\ \citenamefont {Sotiriou}}]{Gumrukcuoglu:2017ijh}%
  \BibitemOpen
  \bibfield  {author} {\bibinfo {author} {\bibfnamefont {A.}~\bibnamefont
  {Emir~Gümrükçüoğlu}}, \bibinfo {author} {\bibfnamefont {M.}~\bibnamefont
  {Saravani}},\ and\ \bibinfo {author} {\bibfnamefont {T.~P.}\ \bibnamefont
  {Sotiriou}},\ }\bibfield  {title} {\bibinfo {title} {{Hořava gravity after
  GW170817}},\ }\href {https://doi.org/10.1103/PhysRevD.97.024032} {\bibfield
  {journal} {\bibinfo  {journal} {Phys. Rev.}\ }\textbf {\bibinfo {volume}
  {D97}},\ \bibinfo {pages} {024032} (\bibinfo {year} {2018})},\ \Eprint
  {https://arxiv.org/abs/1711.08845} {arXiv:1711.08845 [gr-qc]} \BibitemShut
  {NoStop}%
\bibitem [{\citenamefont {Jacobson}\ and\ \citenamefont
  {Mattingly}(2001)}]{Jacobson:2000xp}%
  \BibitemOpen
  \bibfield  {author} {\bibinfo {author} {\bibfnamefont {T.}~\bibnamefont
  {Jacobson}}\ and\ \bibinfo {author} {\bibfnamefont {D.}~\bibnamefont
  {Mattingly}},\ }\bibfield  {title} {\bibinfo {title} {{Gravity with a
  dynamical preferred frame}},\ }\href
  {https://doi.org/10.1103/PhysRevD.64.024028} {\bibfield  {journal} {\bibinfo
  {journal} {Phys. Rev. D}\ }\textbf {\bibinfo {volume} {64}},\ \bibinfo
  {pages} {024028} (\bibinfo {year} {2001})},\ \Eprint
  {https://arxiv.org/abs/gr-qc/0007031} {arXiv:gr-qc/0007031} \BibitemShut
  {NoStop}%
\bibitem [{\citenamefont {Jacobson}(2007)}]{Jacobson:2007veq}%
  \BibitemOpen
  \bibfield  {author} {\bibinfo {author} {\bibfnamefont {T.}~\bibnamefont
  {Jacobson}},\ }\bibfield  {title} {\bibinfo {title} {{Einstein-aether
  gravity: A Status report}},\ }\href {https://doi.org/10.22323/1.043.0020}
  {\bibfield  {journal} {\bibinfo  {journal} {PoS}\ }\textbf {\bibinfo {volume}
  {QG-PH}},\ \bibinfo {pages} {020} (\bibinfo {year} {2007})},\ \Eprint
  {https://arxiv.org/abs/0801.1547} {arXiv:0801.1547 [gr-qc]} \BibitemShut
  {NoStop}%
\bibitem [{\citenamefont {Jacobson}(2014)}]{Jacobson:2013xta}%
  \BibitemOpen
  \bibfield  {author} {\bibinfo {author} {\bibfnamefont {T.}~\bibnamefont
  {Jacobson}},\ }\bibfield  {title} {\bibinfo {title} {{Undoing the twist: The
  Hořava limit of Einstein-aether theory}},\ }\href
  {https://doi.org/10.1103/PhysRevD.89.081501} {\bibfield  {journal} {\bibinfo
  {journal} {Phys. Rev.}\ }\textbf {\bibinfo {volume} {D89}},\ \bibinfo {pages}
  {081501} (\bibinfo {year} {2014})},\ \Eprint
  {https://arxiv.org/abs/1310.5115} {arXiv:1310.5115 [gr-qc]} \BibitemShut
  {NoStop}%
\bibitem [{\citenamefont {Heisenberg}(2014)}]{Heisenberg:2014rta}%
  \BibitemOpen
  \bibfield  {author} {\bibinfo {author} {\bibfnamefont {L.}~\bibnamefont
  {Heisenberg}},\ }\bibfield  {title} {\bibinfo {title} {{Generalization of the
  Proca Action}},\ }\href {https://doi.org/10.1088/1475-7516/2014/05/015}
  {\bibfield  {journal} {\bibinfo  {journal} {JCAP}\ }\textbf {\bibinfo
  {volume} {05}},\ \bibinfo {pages} {015}},\ \Eprint
  {https://arxiv.org/abs/1402.7026} {arXiv:1402.7026 [hep-th]} \BibitemShut
  {NoStop}%
\bibitem [{\citenamefont {Allys}\ \emph
  {et~al.}(2016{\natexlab{a}})\citenamefont {Allys}, \citenamefont {Peter},\
  and\ \citenamefont {Rodriguez}}]{Allys:2015sht}%
  \BibitemOpen
  \bibfield  {author} {\bibinfo {author} {\bibfnamefont {E.}~\bibnamefont
  {Allys}}, \bibinfo {author} {\bibfnamefont {P.}~\bibnamefont {Peter}},\ and\
  \bibinfo {author} {\bibfnamefont {Y.}~\bibnamefont {Rodriguez}},\ }\bibfield
  {title} {\bibinfo {title} {{Generalized Proca action for an Abelian vector
  field}},\ }\href {https://doi.org/10.1088/1475-7516/2016/02/004} {\bibfield
  {journal} {\bibinfo  {journal} {JCAP}\ }\textbf {\bibinfo {volume} {02}},\
  \bibinfo {pages} {004}},\ \Eprint {https://arxiv.org/abs/1511.03101}
  {arXiv:1511.03101 [hep-th]} \BibitemShut {NoStop}%
\bibitem [{\citenamefont {Allys}\ \emph
  {et~al.}(2016{\natexlab{b}})\citenamefont {Allys}, \citenamefont
  {Beltran~Almeida}, \citenamefont {Peter},\ and\ \citenamefont
  {Rodr\'\i{}guez}}]{Allys:2016jaq}%
  \BibitemOpen
  \bibfield  {author} {\bibinfo {author} {\bibfnamefont {E.}~\bibnamefont
  {Allys}}, \bibinfo {author} {\bibfnamefont {J.~P.}\ \bibnamefont
  {Beltran~Almeida}}, \bibinfo {author} {\bibfnamefont {P.}~\bibnamefont
  {Peter}},\ and\ \bibinfo {author} {\bibfnamefont {Y.}~\bibnamefont
  {Rodr\'\i{}guez}},\ }\bibfield  {title} {\bibinfo {title} {{On the 4D
  generalized Proca action for an Abelian vector field}},\ }\href
  {https://doi.org/10.1088/1475-7516/2016/09/026} {\bibfield  {journal}
  {\bibinfo  {journal} {JCAP}\ }\textbf {\bibinfo {volume} {09}},\ \bibinfo
  {pages} {026}},\ \Eprint {https://arxiv.org/abs/1605.08355} {arXiv:1605.08355
  [hep-th]} \BibitemShut {NoStop}%
\bibitem [{\citenamefont {Rodriguez}\ and\ \citenamefont
  {Navarro}(2017)}]{Rodriguez:2017ckc}%
  \BibitemOpen
  \bibfield  {author} {\bibinfo {author} {\bibfnamefont {Y.}~\bibnamefont
  {Rodriguez}}\ and\ \bibinfo {author} {\bibfnamefont {A.~A.}\ \bibnamefont
  {Navarro}},\ }\bibfield  {title} {\bibinfo {title} {{Scalar and vector
  Galileons}},\ }\href {https://doi.org/10.1088/1742-6596/831/1/012004}
  {\bibfield  {journal} {\bibinfo  {journal} {J. Phys. Conf. Ser.}\ }\textbf
  {\bibinfo {volume} {831}},\ \bibinfo {pages} {012004} (\bibinfo {year}
  {2017})},\ \Eprint {https://arxiv.org/abs/1703.01884} {arXiv:1703.01884
  [hep-th]} \BibitemShut {NoStop}%
\bibitem [{\citenamefont {De~Felice}\ \emph {et~al.}(2016)\citenamefont
  {De~Felice}, \citenamefont {Heisenberg}, \citenamefont {Kase}, \citenamefont
  {Tsujikawa}, \citenamefont {Zhang},\ and\ \citenamefont
  {Zhao}}]{DeFelice:2016cri}%
  \BibitemOpen
  \bibfield  {author} {\bibinfo {author} {\bibfnamefont {A.}~\bibnamefont
  {De~Felice}}, \bibinfo {author} {\bibfnamefont {L.}~\bibnamefont
  {Heisenberg}}, \bibinfo {author} {\bibfnamefont {R.}~\bibnamefont {Kase}},
  \bibinfo {author} {\bibfnamefont {S.}~\bibnamefont {Tsujikawa}}, \bibinfo
  {author} {\bibfnamefont {Y.-l.}\ \bibnamefont {Zhang}},\ and\ \bibinfo
  {author} {\bibfnamefont {G.-B.}\ \bibnamefont {Zhao}},\ }\bibfield  {title}
  {\bibinfo {title} {{Screening fifth forces in generalized Proca theories}},\
  }\href {https://doi.org/10.1103/PhysRevD.93.104016} {\bibfield  {journal}
  {\bibinfo  {journal} {Phys. Rev. D}\ }\textbf {\bibinfo {volume} {93}},\
  \bibinfo {pages} {104016} (\bibinfo {year} {2016})},\ \Eprint
  {https://arxiv.org/abs/1602.00371} {arXiv:1602.00371 [gr-qc]} \BibitemShut
  {NoStop}%
\bibitem [{\citenamefont {Gallego~Cadavid}\ \emph {et~al.}(2020)\citenamefont
  {Gallego~Cadavid}, \citenamefont {Rodriguez},\ and\ \citenamefont
  {G\'omez}}]{GallegoCadavid:2020dho}%
  \BibitemOpen
  \bibfield  {author} {\bibinfo {author} {\bibfnamefont {A.}~\bibnamefont
  {Gallego~Cadavid}}, \bibinfo {author} {\bibfnamefont {Y.}~\bibnamefont
  {Rodriguez}},\ and\ \bibinfo {author} {\bibfnamefont {L.~G.}\ \bibnamefont
  {G\'omez}},\ }\bibfield  {title} {\bibinfo {title} {{Generalized SU(2) Proca
  theory reconstructed and beyond}},\ }\href
  {https://doi.org/10.1103/PhysRevD.102.104066} {\bibfield  {journal} {\bibinfo
   {journal} {Phys. Rev. D}\ }\textbf {\bibinfo {volume} {102}},\ \bibinfo
  {pages} {104066} (\bibinfo {year} {2020})},\ \Eprint
  {https://arxiv.org/abs/2009.03241} {arXiv:2009.03241 [hep-th]} \BibitemShut
  {NoStop}%
\bibitem [{\citenamefont {G\'omez}\ and\ \citenamefont
  {Rodr\'\i{}guez}(2019)}]{Gomez:2019tbj}%
  \BibitemOpen
  \bibfield  {author} {\bibinfo {author} {\bibfnamefont {L.~G.}\ \bibnamefont
  {G\'omez}}\ and\ \bibinfo {author} {\bibfnamefont {Y.}~\bibnamefont
  {Rodr\'\i{}guez}},\ }\bibfield  {title} {\bibinfo {title} {{Stability
  Conditions in the Generalized SU(2) Proca Theory}},\ }\href
  {https://doi.org/10.1103/PhysRevD.100.084048} {\bibfield  {journal} {\bibinfo
   {journal} {Phys. Rev. D}\ }\textbf {\bibinfo {volume} {100}},\ \bibinfo
  {pages} {084048} (\bibinfo {year} {2019})},\ \Eprint
  {https://arxiv.org/abs/1907.07961} {arXiv:1907.07961 [gr-qc]} \BibitemShut
  {NoStop}%
\bibitem [{\citenamefont {Nakamura}\ \emph {et~al.}(2019)\citenamefont
  {Nakamura}, \citenamefont {De~Felice}, \citenamefont {Kase},\ and\
  \citenamefont {Tsujikawa}}]{Nakamura:2018oyy}%
  \BibitemOpen
  \bibfield  {author} {\bibinfo {author} {\bibfnamefont {S.}~\bibnamefont
  {Nakamura}}, \bibinfo {author} {\bibfnamefont {A.}~\bibnamefont {De~Felice}},
  \bibinfo {author} {\bibfnamefont {R.}~\bibnamefont {Kase}},\ and\ \bibinfo
  {author} {\bibfnamefont {S.}~\bibnamefont {Tsujikawa}},\ }\bibfield  {title}
  {\bibinfo {title} {{Constraints on massive vector dark energy models from
  integrated Sachs-Wolfe-galaxy cross-correlations}},\ }\href
  {https://doi.org/10.1103/PhysRevD.99.063533} {\bibfield  {journal} {\bibinfo
  {journal} {Phys. Rev. D}\ }\textbf {\bibinfo {volume} {99}},\ \bibinfo
  {pages} {063533} (\bibinfo {year} {2019})},\ \Eprint
  {https://arxiv.org/abs/1811.07541} {arXiv:1811.07541 [astro-ph.CO]}
  \BibitemShut {NoStop}%
\bibitem [{\citenamefont {Heisenberg}\ \emph {et~al.}(2016)\citenamefont
  {Heisenberg}, \citenamefont {Kase},\ and\ \citenamefont
  {Tsujikawa}}]{Heisenberg:2016eld}%
  \BibitemOpen
  \bibfield  {author} {\bibinfo {author} {\bibfnamefont {L.}~\bibnamefont
  {Heisenberg}}, \bibinfo {author} {\bibfnamefont {R.}~\bibnamefont {Kase}},\
  and\ \bibinfo {author} {\bibfnamefont {S.}~\bibnamefont {Tsujikawa}},\
  }\bibfield  {title} {\bibinfo {title} {{Beyond generalized Proca theories}},\
  }\href {https://doi.org/10.1016/j.physletb.2016.07.052} {\bibfield  {journal}
  {\bibinfo  {journal} {Phys. Lett. B}\ }\textbf {\bibinfo {volume} {760}},\
  \bibinfo {pages} {617} (\bibinfo {year} {2016})},\ \Eprint
  {https://arxiv.org/abs/1605.05565} {arXiv:1605.05565 [hep-th]} \BibitemShut
  {NoStop}%
\bibitem [{\citenamefont {Gallego~Cadavid}\ and\ \citenamefont
  {Rodriguez}(2019)}]{GallegoCadavid:2019zke}%
  \BibitemOpen
  \bibfield  {author} {\bibinfo {author} {\bibfnamefont {A.}~\bibnamefont
  {Gallego~Cadavid}}\ and\ \bibinfo {author} {\bibfnamefont {Y.}~\bibnamefont
  {Rodriguez}},\ }\bibfield  {title} {\bibinfo {title} {{A systematic procedure
  to build the beyond generalized Proca field theory}},\ }\href
  {https://doi.org/10.1016/j.physletb.2019.134958} {\bibfield  {journal}
  {\bibinfo  {journal} {Phys. Lett. B}\ }\textbf {\bibinfo {volume} {798}},\
  \bibinfo {pages} {134958} (\bibinfo {year} {2019})},\ \Eprint
  {https://arxiv.org/abs/1905.10664} {arXiv:1905.10664 [hep-th]} \BibitemShut
  {NoStop}%
\bibitem [{\citenamefont {Abbott}\ \emph
  {et~al.}(2017{\natexlab{a}})\citenamefont {Abbott} \emph
  {et~al.}}]{LIGOScientific:2017vwq}%
  \BibitemOpen
  \bibfield  {author} {\bibinfo {author} {\bibfnamefont {B.~P.}\ \bibnamefont
  {Abbott}} \emph {et~al.} (\bibinfo {collaboration} {LIGO Scientific,
  Virgo}),\ }\bibfield  {title} {\bibinfo {title} {{GW170817: Observation of
  Gravitational Waves from a Binary Neutron Star Inspiral}},\ }\href
  {https://doi.org/10.1103/PhysRevLett.119.161101} {\bibfield  {journal}
  {\bibinfo  {journal} {Phys. Rev. Lett.}\ }\textbf {\bibinfo {volume} {119}},\
  \bibinfo {pages} {161101} (\bibinfo {year} {2017}{\natexlab{a}})},\ \Eprint
  {https://arxiv.org/abs/1710.05832} {arXiv:1710.05832 [gr-qc]} \BibitemShut
  {NoStop}%
\bibitem [{\citenamefont {Blas}\ and\ \citenamefont
  {Sanctuary}(2011)}]{Blas:2011zd}%
  \BibitemOpen
  \bibfield  {author} {\bibinfo {author} {\bibfnamefont {D.}~\bibnamefont
  {Blas}}\ and\ \bibinfo {author} {\bibfnamefont {H.}~\bibnamefont
  {Sanctuary}},\ }\bibfield  {title} {\bibinfo {title} {{Gravitational
  Radiation in Ho\v{r}ava Gravity}},\ }\href
  {https://doi.org/10.1103/PhysRevD.84.064004} {\bibfield  {journal} {\bibinfo
  {journal} {Phys. Rev. D}\ }\textbf {\bibinfo {volume} {84}},\ \bibinfo
  {pages} {064004} (\bibinfo {year} {2011})},\ \Eprint
  {https://arxiv.org/abs/1105.5149} {arXiv:1105.5149 [gr-qc]} \BibitemShut
  {NoStop}%
\bibitem [{\citenamefont {Ramos}\ and\ \citenamefont
  {Barausse}(2019)}]{Ramos:2018oku}%
  \BibitemOpen
  \bibfield  {author} {\bibinfo {author} {\bibfnamefont {O.}~\bibnamefont
  {Ramos}}\ and\ \bibinfo {author} {\bibfnamefont {E.}~\bibnamefont
  {Barausse}},\ }\bibfield  {title} {\bibinfo {title} {{Constraints on
  Ho\v{r}ava gravity from binary black hole observations}},\ }\href
  {https://doi.org/10.1103/PhysRevD.99.024034} {\bibfield  {journal} {\bibinfo
  {journal} {Phys. Rev. D}\ }\textbf {\bibinfo {volume} {99}},\ \bibinfo
  {pages} {024034} (\bibinfo {year} {2019})},\ \Eprint
  {https://arxiv.org/abs/1811.07786} {arXiv:1811.07786 [gr-qc]} \BibitemShut
  {NoStop}%
\bibitem [{\citenamefont {Blas}\ \emph {et~al.}(2010)\citenamefont {Blas},
  \citenamefont {Pujolas},\ and\ \citenamefont {Sibiryakov}}]{Blas:2009qj}%
  \BibitemOpen
  \bibfield  {author} {\bibinfo {author} {\bibfnamefont {D.}~\bibnamefont
  {Blas}}, \bibinfo {author} {\bibfnamefont {O.}~\bibnamefont {Pujolas}},\ and\
  \bibinfo {author} {\bibfnamefont {S.}~\bibnamefont {Sibiryakov}},\ }\bibfield
   {title} {\bibinfo {title} {{Consistent Extension of Horava Gravity}},\
  }\href {https://doi.org/10.1103/PhysRevLett.104.181302} {\bibfield  {journal}
  {\bibinfo  {journal} {Phys. Rev. Lett.}\ }\textbf {\bibinfo {volume} {104}},\
  \bibinfo {pages} {181302} (\bibinfo {year} {2010})},\ \Eprint
  {https://arxiv.org/abs/0909.3525} {arXiv:0909.3525 [hep-th]} \BibitemShut
  {NoStop}%
\bibitem [{\citenamefont {Chen}\ \emph {et~al.}(2001)\citenamefont {Chen},
  \citenamefont {Scherrer},\ and\ \citenamefont {Steigman}}]{Chen:2000xxa}%
  \BibitemOpen
  \bibfield  {author} {\bibinfo {author} {\bibfnamefont {X.-l.}\ \bibnamefont
  {Chen}}, \bibinfo {author} {\bibfnamefont {R.~J.}\ \bibnamefont {Scherrer}},\
  and\ \bibinfo {author} {\bibfnamefont {G.}~\bibnamefont {Steigman}},\
  }\bibfield  {title} {\bibinfo {title} {{Extended quintessence and the
  primordial helium abundance}},\ }\href
  {https://doi.org/10.1103/PhysRevD.63.123504} {\bibfield  {journal} {\bibinfo
  {journal} {Phys. Rev. D}\ }\textbf {\bibinfo {volume} {63}},\ \bibinfo
  {pages} {123504} (\bibinfo {year} {2001})},\ \Eprint
  {https://arxiv.org/abs/astro-ph/0011531} {arXiv:astro-ph/0011531}
  \BibitemShut {NoStop}%
\bibitem [{\citenamefont {Carroll}\ and\ \citenamefont
  {Lim}(2004)}]{Carroll:2004ai}%
  \BibitemOpen
  \bibfield  {author} {\bibinfo {author} {\bibfnamefont {S.~M.}\ \bibnamefont
  {Carroll}}\ and\ \bibinfo {author} {\bibfnamefont {E.~A.}\ \bibnamefont
  {Lim}},\ }\bibfield  {title} {\bibinfo {title} {{Lorentz-violating vector
  fields slow the universe down}},\ }\href
  {https://doi.org/10.1103/PhysRevD.70.123525} {\bibfield  {journal} {\bibinfo
  {journal} {Phys. Rev. D}\ }\textbf {\bibinfo {volume} {70}},\ \bibinfo
  {pages} {123525} (\bibinfo {year} {2004})},\ \Eprint
  {https://arxiv.org/abs/hep-th/0407149} {arXiv:hep-th/0407149} \BibitemShut
  {NoStop}%
\bibitem [{\citenamefont {Afshordi}(2009)}]{Afshordi:2009tt}%
  \BibitemOpen
  \bibfield  {author} {\bibinfo {author} {\bibfnamefont {N.}~\bibnamefont
  {Afshordi}},\ }\bibfield  {title} {\bibinfo {title} {{Cuscuton and low energy
  limit of Horava-Lifshitz gravity}},\ }\href
  {https://doi.org/10.1103/PhysRevD.80.081502} {\bibfield  {journal} {\bibinfo
  {journal} {Phys. Rev. D}\ }\textbf {\bibinfo {volume} {80}},\ \bibinfo
  {pages} {081502} (\bibinfo {year} {2009})},\ \Eprint
  {https://arxiv.org/abs/0907.5201} {arXiv:0907.5201 [hep-th]} \BibitemShut
  {NoStop}%
\bibitem [{\citenamefont {Barausse}(2019)}]{Barausse:2019yuk}%
  \BibitemOpen
  \bibfield  {author} {\bibinfo {author} {\bibfnamefont {E.}~\bibnamefont
  {Barausse}},\ }\bibfield  {title} {\bibinfo {title} {{Neutron star
  sensitivities in Ho\v{r}ava gravity after GW170817}},\ }\href
  {https://doi.org/10.1103/PhysRevD.100.084053} {\bibfield  {journal} {\bibinfo
   {journal} {Phys. Rev. D}\ }\textbf {\bibinfo {volume} {100}},\ \bibinfo
  {pages} {084053} (\bibinfo {year} {2019})},\ \Eprint
  {https://arxiv.org/abs/1907.05958} {arXiv:1907.05958 [gr-qc]} \BibitemShut
  {NoStop}%
\bibitem [{\citenamefont {Franchini}\ \emph {et~al.}(2021)\citenamefont
  {Franchini}, \citenamefont {Herrero-Valea},\ and\ \citenamefont
  {Barausse}}]{Franchini:2021bpt}%
  \BibitemOpen
  \bibfield  {author} {\bibinfo {author} {\bibfnamefont {N.}~\bibnamefont
  {Franchini}}, \bibinfo {author} {\bibfnamefont {M.}~\bibnamefont
  {Herrero-Valea}},\ and\ \bibinfo {author} {\bibfnamefont {E.}~\bibnamefont
  {Barausse}},\ }\bibfield  {title} {\bibinfo {title} {{Relation between
  general relativity and a class of Ho\v{r}ava gravity theories}},\ }\href
  {https://doi.org/10.1103/PhysRevD.103.084012} {\bibfield  {journal} {\bibinfo
   {journal} {Phys. Rev. D}\ }\textbf {\bibinfo {volume} {103}},\ \bibinfo
  {pages} {084012} (\bibinfo {year} {2021})},\ \Eprint
  {https://arxiv.org/abs/2103.00929} {arXiv:2103.00929 [gr-qc]} \BibitemShut
  {NoStop}%
\bibitem [{\citenamefont {Yagi}\ and\ \citenamefont
  {Yunes}(2013{\natexlab{a}})}]{Yagi:2013awa}%
  \BibitemOpen
  \bibfield  {author} {\bibinfo {author} {\bibfnamefont {K.}~\bibnamefont
  {Yagi}}\ and\ \bibinfo {author} {\bibfnamefont {N.}~\bibnamefont {Yunes}},\
  }\bibfield  {title} {\bibinfo {title} {{I-Love-Q Relations in Neutron Stars
  and their Applications to Astrophysics, Gravitational Waves and Fundamental
  Physics}},\ }\href {https://doi.org/10.1103/PhysRevD.88.023009} {\bibfield
  {journal} {\bibinfo  {journal} {Phys. Rev. D}\ }\textbf {\bibinfo {volume}
  {88}},\ \bibinfo {pages} {023009} (\bibinfo {year} {2013}{\natexlab{a}})},\
  \Eprint {https://arxiv.org/abs/1303.1528} {arXiv:1303.1528 [gr-qc]}
  \BibitemShut {NoStop}%
\bibitem [{\citenamefont {Yagi}\ and\ \citenamefont
  {Yunes}(2013{\natexlab{b}})}]{Yagi:2013bca}%
  \BibitemOpen
  \bibfield  {author} {\bibinfo {author} {\bibfnamefont {K.}~\bibnamefont
  {Yagi}}\ and\ \bibinfo {author} {\bibfnamefont {N.}~\bibnamefont {Yunes}},\
  }\bibfield  {title} {\bibinfo {title} {{I-Love-Q}},\ }\href
  {https://doi.org/10.1126/science.1236462} {\bibfield  {journal} {\bibinfo
  {journal} {Science}\ }\textbf {\bibinfo {volume} {341}},\ \bibinfo {pages}
  {365} (\bibinfo {year} {2013}{\natexlab{b}})},\ \Eprint
  {https://arxiv.org/abs/1302.4499} {arXiv:1302.4499 [gr-qc]} \BibitemShut
  {NoStop}%
\bibitem [{\citenamefont {Yagi}\ and\ \citenamefont
  {Yunes}(2017)}]{Yagi:2016bkt}%
  \BibitemOpen
  \bibfield  {author} {\bibinfo {author} {\bibfnamefont {K.}~\bibnamefont
  {Yagi}}\ and\ \bibinfo {author} {\bibfnamefont {N.}~\bibnamefont {Yunes}},\
  }\bibfield  {title} {\bibinfo {title} {{Approximate Universal Relations for
  Neutron Stars and Quark Stars}},\ }\href
  {https://doi.org/10.1016/j.physrep.2017.03.002} {\bibfield  {journal}
  {\bibinfo  {journal} {Phys. Rept.}\ }\textbf {\bibinfo {volume} {681}},\
  \bibinfo {pages} {1} (\bibinfo {year} {2017})},\ \Eprint
  {https://arxiv.org/abs/1608.02582} {arXiv:1608.02582 [gr-qc]} \BibitemShut
  {NoStop}%
\bibitem [{\citenamefont {Doneva}\ and\ \citenamefont
  {Pappas}(2018)}]{Doneva:2017jop}%
  \BibitemOpen
  \bibfield  {author} {\bibinfo {author} {\bibfnamefont {D.~D.}\ \bibnamefont
  {Doneva}}\ and\ \bibinfo {author} {\bibfnamefont {G.}~\bibnamefont
  {Pappas}},\ }\bibfield  {title} {\bibinfo {title} {{Universal Relations and
  Alternative Gravity Theories}},\ }\href
  {https://doi.org/10.1007/978-3-319-97616-7_13} {\bibfield  {journal}
  {\bibinfo  {journal} {Astrophys. Space Sci. Libr.}\ }\textbf {\bibinfo
  {volume} {457}},\ \bibinfo {pages} {737} (\bibinfo {year} {2018})},\ \Eprint
  {https://arxiv.org/abs/1709.08046} {arXiv:1709.08046 [gr-qc]} \BibitemShut
  {NoStop}%
\bibitem [{\citenamefont {Gupta}\ \emph {et~al.}(2018)\citenamefont {Gupta},
  \citenamefont {Majumder}, \citenamefont {Yagi},\ and\ \citenamefont
  {Yunes}}]{Gupta:2017vsl}%
  \BibitemOpen
  \bibfield  {author} {\bibinfo {author} {\bibfnamefont {T.}~\bibnamefont
  {Gupta}}, \bibinfo {author} {\bibfnamefont {B.}~\bibnamefont {Majumder}},
  \bibinfo {author} {\bibfnamefont {K.}~\bibnamefont {Yagi}},\ and\ \bibinfo
  {author} {\bibfnamefont {N.}~\bibnamefont {Yunes}},\ }\bibfield  {title}
  {\bibinfo {title} {{I-Love-Q Relations for Neutron Stars in dynamical Chern
  Simons Gravity}},\ }\href {https://doi.org/10.1088/1361-6382/aa9c68}
  {\bibfield  {journal} {\bibinfo  {journal} {Class. Quant. Grav.}\ }\textbf
  {\bibinfo {volume} {35}},\ \bibinfo {pages} {025009} (\bibinfo {year}
  {2018})},\ \Eprint {https://arxiv.org/abs/1710.07862} {arXiv:1710.07862
  [gr-qc]} \BibitemShut {NoStop}%
\bibitem [{\citenamefont {Abbott}\ \emph {et~al.}(2019)\citenamefont {Abbott}
  \emph {et~al.}}]{LIGOScientific:2018hze}%
  \BibitemOpen
  \bibfield  {author} {\bibinfo {author} {\bibfnamefont {B.~P.}\ \bibnamefont
  {Abbott}} \emph {et~al.} (\bibinfo {collaboration} {LIGO Scientific,
  Virgo}),\ }\bibfield  {title} {\bibinfo {title} {{Properties of the binary
  neutron star merger GW170817}},\ }\href
  {https://doi.org/10.1103/PhysRevX.9.011001} {\bibfield  {journal} {\bibinfo
  {journal} {Phys. Rev. X}\ }\textbf {\bibinfo {volume} {9}},\ \bibinfo {pages}
  {011001} (\bibinfo {year} {2019})},\ \Eprint
  {https://arxiv.org/abs/1805.11579} {arXiv:1805.11579 [gr-qc]} \BibitemShut
  {NoStop}%
\bibitem [{\citenamefont {Abbott}\ \emph {et~al.}(2018)\citenamefont {Abbott}
  \emph {et~al.}}]{LIGOScientific:2018cki}%
  \BibitemOpen
  \bibfield  {author} {\bibinfo {author} {\bibfnamefont {B.~P.}\ \bibnamefont
  {Abbott}} \emph {et~al.} (\bibinfo {collaboration} {LIGO Scientific,
  Virgo}),\ }\bibfield  {title} {\bibinfo {title} {{GW170817: Measurements of
  neutron star radii and equation of state}},\ }\href
  {https://doi.org/10.1103/PhysRevLett.121.161101} {\bibfield  {journal}
  {\bibinfo  {journal} {Phys. Rev. Lett.}\ }\textbf {\bibinfo {volume} {121}},\
  \bibinfo {pages} {161101} (\bibinfo {year} {2018})},\ \Eprint
  {https://arxiv.org/abs/1805.11581} {arXiv:1805.11581 [gr-qc]} \BibitemShut
  {NoStop}%
\bibitem [{\citenamefont {Silva}\ \emph {et~al.}(2021)\citenamefont {Silva},
  \citenamefont {Holgado}, \citenamefont {C\'ardenas-Avenda\~no},\ and\
  \citenamefont {Yunes}}]{Silva:2020acr}%
  \BibitemOpen
  \bibfield  {author} {\bibinfo {author} {\bibfnamefont {H.~O.}\ \bibnamefont
  {Silva}}, \bibinfo {author} {\bibfnamefont {A.~M.}\ \bibnamefont {Holgado}},
  \bibinfo {author} {\bibfnamefont {A.}~\bibnamefont {C\'ardenas-Avenda\~no}},\
  and\ \bibinfo {author} {\bibfnamefont {N.}~\bibnamefont {Yunes}},\ }\bibfield
   {title} {\bibinfo {title} {{Astrophysical and theoretical physics
  implications from multimessenger neutron star observations}},\ }\href
  {https://doi.org/10.1103/PhysRevLett.126.181101} {\bibfield  {journal}
  {\bibinfo  {journal} {Phys. Rev. Lett.}\ }\textbf {\bibinfo {volume} {126}},\
  \bibinfo {pages} {181101} (\bibinfo {year} {2021})},\ \Eprint
  {https://arxiv.org/abs/2004.01253} {arXiv:2004.01253 [gr-qc]} \BibitemShut
  {NoStop}%
\bibitem [{\citenamefont {Lattimer}\ and\ \citenamefont
  {Schutz}(2005)}]{Lattimer:2004nj}%
  \BibitemOpen
  \bibfield  {author} {\bibinfo {author} {\bibfnamefont {J.~M.}\ \bibnamefont
  {Lattimer}}\ and\ \bibinfo {author} {\bibfnamefont {B.~F.}\ \bibnamefont
  {Schutz}},\ }\bibfield  {title} {\bibinfo {title} {{Constraining the equation
  of state with moment of inertia measurements}},\ }\href
  {https://doi.org/10.1086/431543} {\bibfield  {journal} {\bibinfo  {journal}
  {Astrophys. J.}\ }\textbf {\bibinfo {volume} {629}},\ \bibinfo {pages} {979}
  (\bibinfo {year} {2005})},\ \Eprint {https://arxiv.org/abs/astro-ph/0411470}
  {arXiv:astro-ph/0411470} \BibitemShut {NoStop}%
\bibitem [{\citenamefont {Kramer}\ and\ \citenamefont
  {Wex}(2009)}]{Kramer:2009zza}%
  \BibitemOpen
  \bibfield  {author} {\bibinfo {author} {\bibfnamefont {M.}~\bibnamefont
  {Kramer}}\ and\ \bibinfo {author} {\bibfnamefont {N.}~\bibnamefont {Wex}},\
  }\bibfield  {title} {\bibinfo {title} {{The double pulsar system: A unique
  laboratory for gravity}},\ }\href
  {https://doi.org/10.1088/0264-9381/26/7/073001} {\bibfield  {journal}
  {\bibinfo  {journal} {Class. Quant. Grav.}\ }\textbf {\bibinfo {volume}
  {26}},\ \bibinfo {pages} {073001} (\bibinfo {year} {2009})}\BibitemShut
  {NoStop}%
\bibitem [{\citenamefont {Hu}\ \emph {et~al.}(2020)\citenamefont {Hu},
  \citenamefont {Kramer}, \citenamefont {Wex}, \citenamefont {Champion},\ and\
  \citenamefont {Kehl}}]{Hu:2020ubl}%
  \BibitemOpen
  \bibfield  {author} {\bibinfo {author} {\bibfnamefont {H.}~\bibnamefont
  {Hu}}, \bibinfo {author} {\bibfnamefont {M.}~\bibnamefont {Kramer}}, \bibinfo
  {author} {\bibfnamefont {N.}~\bibnamefont {Wex}}, \bibinfo {author}
  {\bibfnamefont {D.~J.}\ \bibnamefont {Champion}},\ and\ \bibinfo {author}
  {\bibfnamefont {M.~S.}\ \bibnamefont {Kehl}},\ }\bibfield  {title} {\bibinfo
  {title} {{Constraining the dense matter equation-of-state with radio
  pulsars}},\ }\href {https://doi.org/10.1093/mnras/staa2107} {\bibfield
  {journal} {\bibinfo  {journal} {Mon. Not. Roy. Astron. Soc.}\ }\textbf
  {\bibinfo {volume} {497}},\ \bibinfo {pages} {3118} (\bibinfo {year}
  {2020})},\ \Eprint {https://arxiv.org/abs/2007.07725} {arXiv:2007.07725
  [astro-ph.SR]} \BibitemShut {NoStop}%
\bibitem [{\citenamefont {Miller}\ \emph {et~al.}(2019)\citenamefont {Miller}
  \emph {et~al.}}]{Miller:2019cac}%
  \BibitemOpen
  \bibfield  {author} {\bibinfo {author} {\bibfnamefont {M.~C.}\ \bibnamefont
  {Miller}} \emph {et~al.},\ }\bibfield  {title} {\bibinfo {title} {{PSR
  J0030+0451 Mass and Radius from $NICER$ Data and Implications for the
  Properties of Neutron Star Matter}},\ }\href
  {https://doi.org/10.3847/2041-8213/ab50c5} {\bibfield  {journal} {\bibinfo
  {journal} {Astrophys. J. Lett.}\ }\textbf {\bibinfo {volume} {887}},\
  \bibinfo {pages} {L24} (\bibinfo {year} {2019})},\ \Eprint
  {https://arxiv.org/abs/1912.05705} {arXiv:1912.05705 [astro-ph.HE]}
  \BibitemShut {NoStop}%
\bibitem [{\citenamefont {Riley}\ \emph {et~al.}(2019)\citenamefont {Riley}
  \emph {et~al.}}]{Riley:2019yda}%
  \BibitemOpen
  \bibfield  {author} {\bibinfo {author} {\bibfnamefont {T.~E.}\ \bibnamefont
  {Riley}} \emph {et~al.},\ }\bibfield  {title} {\bibinfo {title} {{A $NICER$
  View of PSR J0030+0451: Millisecond Pulsar Parameter Estimation}},\ }\href
  {https://doi.org/10.3847/2041-8213/ab481c} {\bibfield  {journal} {\bibinfo
  {journal} {Astrophys. J. Lett.}\ }\textbf {\bibinfo {volume} {887}},\
  \bibinfo {pages} {L21} (\bibinfo {year} {2019})},\ \Eprint
  {https://arxiv.org/abs/1912.05702} {arXiv:1912.05702 [astro-ph.HE]}
  \BibitemShut {NoStop}%
\bibitem [{\citenamefont {Saffer}\ and\ \citenamefont
  {Yagi}(2021)}]{Saffer:2021gak}%
  \BibitemOpen
  \bibfield  {author} {\bibinfo {author} {\bibfnamefont {A.}~\bibnamefont
  {Saffer}}\ and\ \bibinfo {author} {\bibfnamefont {K.}~\bibnamefont {Yagi}},\
  }\bibfield  {title} {\bibinfo {title} {{Tidal deformabilities of neutron
  stars in scalar-Gauss-Bonnet gravity and their applications to multimessenger
  tests of gravity}},\ }\href {https://doi.org/10.1103/PhysRevD.104.124052}
  {\bibfield  {journal} {\bibinfo  {journal} {Phys. Rev. D}\ }\textbf {\bibinfo
  {volume} {104}},\ \bibinfo {pages} {124052} (\bibinfo {year} {2021})},\
  \Eprint {https://arxiv.org/abs/2110.02997} {arXiv:2110.02997 [gr-qc]}
  \BibitemShut {NoStop}%
\bibitem [{\citenamefont {Bellorin}\ and\ \citenamefont
  {Restuccia}(2012)}]{Bellorin:2010je}%
  \BibitemOpen
  \bibfield  {author} {\bibinfo {author} {\bibfnamefont {J.}~\bibnamefont
  {Bellorin}}\ and\ \bibinfo {author} {\bibfnamefont {A.}~\bibnamefont
  {Restuccia}},\ }\bibfield  {title} {\bibinfo {title} {{On the consistency of
  the Horava Theory}},\ }\href {https://doi.org/10.1142/S021827182500290}
  {\bibfield  {journal} {\bibinfo  {journal} {Int. J. Mod. Phys. D}\ }\textbf
  {\bibinfo {volume} {21}},\ \bibinfo {pages} {1250029} (\bibinfo {year}
  {2012})},\ \Eprint {https://arxiv.org/abs/1004.0055} {arXiv:1004.0055
  [hep-th]} \BibitemShut {NoStop}%
\bibitem [{\citenamefont {Gupta}\ \emph {et~al.}(2021)\citenamefont {Gupta},
  \citenamefont {Herrero-Valea}, \citenamefont {Blas}, \citenamefont
  {Barausse}, \citenamefont {Cornish}, \citenamefont {Yagi},\ and\
  \citenamefont {Yunes}}]{Gupta:2021vdj}%
  \BibitemOpen
  \bibfield  {author} {\bibinfo {author} {\bibfnamefont {T.}~\bibnamefont
  {Gupta}}, \bibinfo {author} {\bibfnamefont {M.}~\bibnamefont
  {Herrero-Valea}}, \bibinfo {author} {\bibfnamefont {D.}~\bibnamefont {Blas}},
  \bibinfo {author} {\bibfnamefont {E.}~\bibnamefont {Barausse}}, \bibinfo
  {author} {\bibfnamefont {N.}~\bibnamefont {Cornish}}, \bibinfo {author}
  {\bibfnamefont {K.}~\bibnamefont {Yagi}},\ and\ \bibinfo {author}
  {\bibfnamefont {N.}~\bibnamefont {Yunes}},\ }\bibfield  {title} {\bibinfo
  {title} {{New binary pulsar constraints on Einstein-\ae{}ther theory after
  GW170817}},\ }\href {https://doi.org/10.1088/1361-6382/ac1a69} {\bibfield
  {journal} {\bibinfo  {journal} {Class. Quant. Grav.}\ }\textbf {\bibinfo
  {volume} {38}},\ \bibinfo {pages} {195003} (\bibinfo {year} {2021})},\
  \Eprint {https://arxiv.org/abs/2104.04596} {arXiv:2104.04596 [gr-qc]}
  \BibitemShut {NoStop}%
\bibitem [{\citenamefont {Yagi}\ and\ \citenamefont
  {Stepniczka}(2021)}]{Yagi:2021loe}%
  \BibitemOpen
  \bibfield  {author} {\bibinfo {author} {\bibfnamefont {K.}~\bibnamefont
  {Yagi}}\ and\ \bibinfo {author} {\bibfnamefont {M.}~\bibnamefont
  {Stepniczka}},\ }\bibfield  {title} {\bibinfo {title} {{Neutron stars in
  scalar-tensor theories: Analytic scalar charges and universal relations}},\
  }\href {https://doi.org/10.1103/PhysRevD.104.044017} {\bibfield  {journal}
  {\bibinfo  {journal} {Phys. Rev. D}\ }\textbf {\bibinfo {volume} {104}},\
  \bibinfo {pages} {044017} (\bibinfo {year} {2021})},\ \Eprint
  {https://arxiv.org/abs/2105.01614} {arXiv:2105.01614 [gr-qc]} \BibitemShut
  {NoStop}%
\bibitem [{\citenamefont {Pere\~niguez}\ and\ \citenamefont
  {Cardoso}(2022)}]{Pereniguez:2021xcj}%
  \BibitemOpen
  \bibfield  {author} {\bibinfo {author} {\bibfnamefont {D.}~\bibnamefont
  {Pere\~niguez}}\ and\ \bibinfo {author} {\bibfnamefont {V.}~\bibnamefont
  {Cardoso}},\ }\bibfield  {title} {\bibinfo {title} {{Love numbers and
  magnetic susceptibility of charged black holes}},\ }\href
  {https://doi.org/10.1103/PhysRevD.105.044026} {\bibfield  {journal} {\bibinfo
   {journal} {Phys. Rev. D}\ }\textbf {\bibinfo {volume} {105}},\ \bibinfo
  {pages} {044026} (\bibinfo {year} {2022})},\ \Eprint
  {https://arxiv.org/abs/2112.08400} {arXiv:2112.08400 [gr-qc]} \BibitemShut
  {NoStop}%
\bibitem [{\citenamefont {Poisson}(2009)}]{Poisson:2009pwt}%
  \BibitemOpen
  \bibfield  {author} {\bibinfo {author} {\bibfnamefont {E.}~\bibnamefont
  {Poisson}},\ }\href {https://doi.org/10.1017/CBO9780511606601} {\emph
  {\bibinfo {title} {{A Relativist's Toolkit: The Mathematics of Black-Hole
  Mechanics}}}}\ (\bibinfo  {publisher} {Cambridge University Press},\ \bibinfo
  {year} {2009})\BibitemShut {NoStop}%
\bibitem [{\citenamefont {Abbott}\ \emph
  {et~al.}(2017{\natexlab{b}})\citenamefont {Abbott} \emph
  {et~al.}}]{LIGOScientific:2017zic}%
  \BibitemOpen
  \bibfield  {author} {\bibinfo {author} {\bibfnamefont {B.~P.}\ \bibnamefont
  {Abbott}} \emph {et~al.} (\bibinfo {collaboration} {LIGO Scientific, Virgo,
  Fermi-GBM, INTEGRAL}),\ }\bibfield  {title} {\bibinfo {title} {{Gravitational
  Waves and Gamma-rays from a Binary Neutron Star Merger: GW170817 and GRB
  170817A}},\ }\href {https://doi.org/10.3847/2041-8213/aa920c} {\bibfield
  {journal} {\bibinfo  {journal} {Astrophys. J. Lett.}\ }\textbf {\bibinfo
  {volume} {848}},\ \bibinfo {pages} {L13} (\bibinfo {year}
  {2017}{\natexlab{b}})},\ \Eprint {https://arxiv.org/abs/1710.05834}
  {arXiv:1710.05834 [astro-ph.HE]} \BibitemShut {NoStop}%
\bibitem [{\citenamefont {Regge}\ and\ \citenamefont
  {Wheeler}(1957)}]{Regge:1957td}%
  \BibitemOpen
  \bibfield  {author} {\bibinfo {author} {\bibfnamefont {T.}~\bibnamefont
  {Regge}}\ and\ \bibinfo {author} {\bibfnamefont {J.~A.}\ \bibnamefont
  {Wheeler}},\ }\bibfield  {title} {\bibinfo {title} {{Stability of a
  Schwarzschild singularity}},\ }\href
  {https://doi.org/10.1103/PhysRev.108.1063} {\bibfield  {journal} {\bibinfo
  {journal} {Phys. Rev.}\ }\textbf {\bibinfo {volume} {108}},\ \bibinfo {pages}
  {1063} (\bibinfo {year} {1957})}\BibitemShut {NoStop}%
\bibitem [{\citenamefont {{Thorne}}\ and\ \citenamefont
  {{Campolattaro}}(1967)}]{Thorne:1967}%
  \BibitemOpen
  \bibfield  {author} {\bibinfo {author} {\bibfnamefont {K.~S.}\ \bibnamefont
  {{Thorne}}}\ and\ \bibinfo {author} {\bibfnamefont {A.}~\bibnamefont
  {{Campolattaro}}},\ }\bibfield  {title} {\bibinfo {title} {{Non-Radial
  Pulsation of General-Relativistic Stellar Models. I. Analytic Analysis for $l
  \geq $ 2}},\ }\href {https://doi.org/10.1086/149288} {\bibfield  {journal}
  {\bibinfo  {journal} {\apj}\ }\textbf {\bibinfo {volume} {149}},\ \bibinfo
  {pages} {591} (\bibinfo {year} {1967})}\BibitemShut {NoStop}%
\bibitem [{\citenamefont {Pani}\ \emph {et~al.}(2013)\citenamefont {Pani},
  \citenamefont {Berti},\ and\ \citenamefont {Gualtieri}}]{Pani:2013wsa}%
  \BibitemOpen
  \bibfield  {author} {\bibinfo {author} {\bibfnamefont {P.}~\bibnamefont
  {Pani}}, \bibinfo {author} {\bibfnamefont {E.}~\bibnamefont {Berti}},\ and\
  \bibinfo {author} {\bibfnamefont {L.}~\bibnamefont {Gualtieri}},\ }\bibfield
  {title} {\bibinfo {title} {{Scalar, Electromagnetic and Gravitational
  Perturbations of Kerr-Newman Black Holes in the Slow-Rotation Limit}},\
  }\href {https://doi.org/10.1103/PhysRevD.88.064048} {\bibfield  {journal}
  {\bibinfo  {journal} {Phys. Rev. D}\ }\textbf {\bibinfo {volume} {88}},\
  \bibinfo {pages} {064048} (\bibinfo {year} {2013})},\ \Eprint
  {https://arxiv.org/abs/1307.7315} {arXiv:1307.7315 [gr-qc]} \BibitemShut
  {NoStop}%
\bibitem [{\citenamefont {Friedman}\ and\ \citenamefont
  {Stergioulas}(2013)}]{Friedman:2013xza}%
  \BibitemOpen
  \bibfield  {author} {\bibinfo {author} {\bibfnamefont {J.~L.}\ \bibnamefont
  {Friedman}}\ and\ \bibinfo {author} {\bibfnamefont {N.}~\bibnamefont
  {Stergioulas}},\ }\href {https://doi.org/10.1017/CBO9780511977596} {\emph
  {\bibinfo {title} {{Rotating Relativistic Stars}}}},\ Cambridge Monographs on
  Mathematical Physics\ (\bibinfo  {publisher} {Cambridge University Press},\
  \bibinfo {year} {2013})\BibitemShut {NoStop}%
\bibitem [{\citenamefont {Pani}\ and\ \citenamefont
  {Berti}(2014)}]{Pani:2014jra}%
  \BibitemOpen
  \bibfield  {author} {\bibinfo {author} {\bibfnamefont {P.}~\bibnamefont
  {Pani}}\ and\ \bibinfo {author} {\bibfnamefont {E.}~\bibnamefont {Berti}},\
  }\bibfield  {title} {\bibinfo {title} {{Slowly rotating neutron stars in
  scalar-tensor theories}},\ }\href
  {https://doi.org/10.1103/PhysRevD.90.024025} {\bibfield  {journal} {\bibinfo
  {journal} {Phys. Rev. D}\ }\textbf {\bibinfo {volume} {90}},\ \bibinfo
  {pages} {024025} (\bibinfo {year} {2014})},\ \Eprint
  {https://arxiv.org/abs/1405.4547} {arXiv:1405.4547 [gr-qc]} \BibitemShut
  {NoStop}%
\bibitem [{\citenamefont {Cardoso}\ \emph {et~al.}(2017)\citenamefont
  {Cardoso}, \citenamefont {Franzin}, \citenamefont {Maselli}, \citenamefont
  {Pani},\ and\ \citenamefont {Raposo}}]{Cardoso:2017cfl}%
  \BibitemOpen
  \bibfield  {author} {\bibinfo {author} {\bibfnamefont {V.}~\bibnamefont
  {Cardoso}}, \bibinfo {author} {\bibfnamefont {E.}~\bibnamefont {Franzin}},
  \bibinfo {author} {\bibfnamefont {A.}~\bibnamefont {Maselli}}, \bibinfo
  {author} {\bibfnamefont {P.}~\bibnamefont {Pani}},\ and\ \bibinfo {author}
  {\bibfnamefont {G.}~\bibnamefont {Raposo}},\ }\bibfield  {title} {\bibinfo
  {title} {{Testing strong-field gravity with tidal Love numbers}},\ }\href
  {https://doi.org/10.1103/PhysRevD.95.084014} {\bibfield  {journal} {\bibinfo
  {journal} {Phys. Rev. D}\ }\textbf {\bibinfo {volume} {95}},\ \bibinfo
  {pages} {084014} (\bibinfo {year} {2017})},\ \bibinfo {note} {[Addendum:
  Phys.Rev.D 95, 089901 (2017)]},\ \Eprint {https://arxiv.org/abs/1701.01116}
  {arXiv:1701.01116 [gr-qc]} \BibitemShut {NoStop}%
\bibitem [{\citenamefont {Tolman}(1939)}]{Tolman:1939jz}%
  \BibitemOpen
  \bibfield  {author} {\bibinfo {author} {\bibfnamefont {R.~C.}\ \bibnamefont
  {Tolman}},\ }\bibfield  {title} {\bibinfo {title} {{Static solutions of
  Einstein's field equations for spheres of fluid}},\ }\href
  {https://doi.org/10.1103/PhysRev.55.364} {\bibfield  {journal} {\bibinfo
  {journal} {Phys. Rev.}\ }\textbf {\bibinfo {volume} {55}},\ \bibinfo {pages}
  {364} (\bibinfo {year} {1939})}\BibitemShut {NoStop}%
\bibitem [{\citenamefont {Lattimer}\ and\ \citenamefont
  {Prakash}(2001)}]{Lattimer:2000nx}%
  \BibitemOpen
  \bibfield  {author} {\bibinfo {author} {\bibfnamefont {J.~M.}\ \bibnamefont
  {Lattimer}}\ and\ \bibinfo {author} {\bibfnamefont {M.}~\bibnamefont
  {Prakash}},\ }\bibfield  {title} {\bibinfo {title} {{Neutron star structure
  and the equation of state}},\ }\href {https://doi.org/10.1086/319702}
  {\bibfield  {journal} {\bibinfo  {journal} {Astrophys. J.}\ }\textbf
  {\bibinfo {volume} {550}},\ \bibinfo {pages} {426} (\bibinfo {year}
  {2001})},\ \Eprint {https://arxiv.org/abs/astro-ph/0002232}
  {arXiv:astro-ph/0002232} \BibitemShut {NoStop}%
\bibitem [{\citenamefont {Jiang}\ and\ \citenamefont
  {Yagi}(2019)}]{Jiang:2019vmf}%
  \BibitemOpen
  \bibfield  {author} {\bibinfo {author} {\bibfnamefont {N.}~\bibnamefont
  {Jiang}}\ and\ \bibinfo {author} {\bibfnamefont {K.}~\bibnamefont {Yagi}},\
  }\bibfield  {title} {\bibinfo {title} {{Improved Analytic Modeling of Neutron
  Star Interiors}},\ }\href {https://doi.org/10.1103/PhysRevD.99.124029}
  {\bibfield  {journal} {\bibinfo  {journal} {Phys. Rev. D}\ }\textbf {\bibinfo
  {volume} {99}},\ \bibinfo {pages} {124029} (\bibinfo {year} {2019})},\
  \Eprint {https://arxiv.org/abs/1904.05954} {arXiv:1904.05954 [gr-qc]}
  \BibitemShut {NoStop}%
\bibitem [{\citenamefont {Yagi}(2014)}]{Yagi:2013sva}%
  \BibitemOpen
  \bibfield  {author} {\bibinfo {author} {\bibfnamefont {K.}~\bibnamefont
  {Yagi}},\ }\bibfield  {title} {\bibinfo {title} {{Multipole Love
  Relations}},\ }\href {https://doi.org/10.1103/PhysRevD.89.043011} {\bibfield
  {journal} {\bibinfo  {journal} {Phys. Rev. D}\ }\textbf {\bibinfo {volume}
  {89}},\ \bibinfo {pages} {043011} (\bibinfo {year} {2014})},\ \bibinfo {note}
  {[Erratum: Phys.Rev.D 96, 129904 (2017), Erratum: Phys.Rev.D 97, 129901
  (2018)]},\ \Eprint {https://arxiv.org/abs/1311.0872} {arXiv:1311.0872
  [gr-qc]} \BibitemShut {NoStop}%
\bibitem [{\citenamefont {Gralla}(2018)}]{Gralla:2017djj}%
  \BibitemOpen
  \bibfield  {author} {\bibinfo {author} {\bibfnamefont {S.~E.}\ \bibnamefont
  {Gralla}},\ }\bibfield  {title} {\bibinfo {title} {{On the Ambiguity in
  Relativistic Tidal Deformability}},\ }\href
  {https://doi.org/10.1088/1361-6382/aab186} {\bibfield  {journal} {\bibinfo
  {journal} {Class. Quant. Grav.}\ }\textbf {\bibinfo {volume} {35}},\ \bibinfo
  {pages} {085002} (\bibinfo {year} {2018})},\ \Eprint
  {https://arxiv.org/abs/1710.11096} {arXiv:1710.11096 [gr-qc]} \BibitemShut
  {NoStop}%
\bibitem [{\citenamefont {Creci}\ \emph {et~al.}(2021)\citenamefont {Creci},
  \citenamefont {Hinderer},\ and\ \citenamefont {Steinhoff}}]{Creci:2021rkz}%
  \BibitemOpen
  \bibfield  {author} {\bibinfo {author} {\bibfnamefont {G.}~\bibnamefont
  {Creci}}, \bibinfo {author} {\bibfnamefont {T.}~\bibnamefont {Hinderer}},\
  and\ \bibinfo {author} {\bibfnamefont {J.}~\bibnamefont {Steinhoff}},\
  }\bibfield  {title} {\bibinfo {title} {{Tidal response from scattering and
  the role of analytic continuation}},\ }\href
  {https://doi.org/10.1103/PhysRevD.104.124061} {\bibfield  {journal} {\bibinfo
   {journal} {Phys. Rev. D}\ }\textbf {\bibinfo {volume} {104}},\ \bibinfo
  {pages} {124061} (\bibinfo {year} {2021})},\ \Eprint
  {https://arxiv.org/abs/2108.03385} {arXiv:2108.03385 [gr-qc]} \BibitemShut
  {NoStop}%
\bibitem [{\citenamefont {Remie}\ and\ \citenamefont {Bonga}(2021)}]{remie}%
  \BibitemOpen
  \bibfield  {author} {\bibinfo {author} {\bibfnamefont {J.}~\bibnamefont
  {Remie}}\ and\ \bibinfo {author} {\bibfnamefont {B.~P.}\ \bibnamefont
  {Bonga}},\ }\href@noop {} {\bibinfo {title} {Calculating $\mathrm{L}$ove
  numbers using black hole perturbation theory (master thesis)}} (\bibinfo
  {year} {2021})\BibitemShut {NoStop}%
\bibitem [{\citenamefont {Zhu}\ \emph {et~al.}(2020)\citenamefont {Zhu},
  \citenamefont {Li},\ and\ \citenamefont {Rezzolla}}]{Zhu:2020imp}%
  \BibitemOpen
  \bibfield  {author} {\bibinfo {author} {\bibfnamefont {Z.}~\bibnamefont
  {Zhu}}, \bibinfo {author} {\bibfnamefont {A.}~\bibnamefont {Li}},\ and\
  \bibinfo {author} {\bibfnamefont {L.}~\bibnamefont {Rezzolla}},\ }\bibfield
  {title} {\bibinfo {title} {{Tidal deformability and gravitational-wave phase
  evolution of magnetized compact-star binaries}},\ }\href
  {https://doi.org/10.1103/PhysRevD.102.084058} {\bibfield  {journal} {\bibinfo
   {journal} {Phys. Rev. D}\ }\textbf {\bibinfo {volume} {102}},\ \bibinfo
  {pages} {084058} (\bibinfo {year} {2020})},\ \Eprint
  {https://arxiv.org/abs/2005.02677} {arXiv:2005.02677 [astro-ph.HE]}
  \BibitemShut {NoStop}%
\bibitem [{\citenamefont {Vylet}\ \emph {et~al.}(tion)\citenamefont {Vylet},
  \citenamefont {Ajith}, \citenamefont {Yagi},\ and\ \citenamefont
  {Yunes}}]{kai}%
  \BibitemOpen
  \bibfield  {author} {\bibinfo {author} {\bibfnamefont {K.}~\bibnamefont
  {Vylet}}, \bibinfo {author} {\bibfnamefont {S.}~\bibnamefont {Ajith}},
  \bibinfo {author} {\bibfnamefont {K.}~\bibnamefont {Yagi}},\ and\ \bibinfo
  {author} {\bibfnamefont {N.}~\bibnamefont {Yunes}},\ }\href@noop {} {}
  (\bibinfo {year} {in preparation})\BibitemShut {NoStop}%
\bibitem [{\citenamefont {Barausse}\ and\ \citenamefont
  {Sotiriou}(2013)}]{Barausse:2012qh}%
  \BibitemOpen
  \bibfield  {author} {\bibinfo {author} {\bibfnamefont {E.}~\bibnamefont
  {Barausse}}\ and\ \bibinfo {author} {\bibfnamefont {T.~P.}\ \bibnamefont
  {Sotiriou}},\ }\bibfield  {title} {\bibinfo {title} {{Slowly rotating black
  holes in Horava-Lifshitz gravity}},\ }\href
  {https://doi.org/10.1103/PhysRevD.87.087504} {\bibfield  {journal} {\bibinfo
  {journal} {Phys. Rev. D}\ }\textbf {\bibinfo {volume} {87}},\ \bibinfo
  {pages} {087504} (\bibinfo {year} {2013})},\ \Eprint
  {https://arxiv.org/abs/1212.1334} {arXiv:1212.1334 [gr-qc]} \BibitemShut
  {NoStop}%
\end{thebibliography}%
\end{document}